\documentclass[twocolumn,english,amsmath,amssymb,aps,reprint,pra]{revtex4-2}
\usepackage[T1]{fontenc}
\usepackage{xcolor}
\usepackage{amsmath}
\usepackage{amssymb}
\usepackage[pdfusetitle,
 bookmarks=true,bookmarksnumbered=false,bookmarksopen=false,
 breaklinks=false,pdfborder={0 0 0},pdfborderstyle={},backref=false,colorlinks=true]
 {hyperref}
\hypersetup{
 colorlinks=true,linkcolor=blue,citecolor=blue,urlcolor=blue}

\makeatletter
\usepackage{graphicx}% Include figure files
\usepackage{braket}
\usepackage{multirow}
\usepackage{tikz}
\usetikzlibrary{arrows.meta,calc,positioning,decorations.pathmorphing}

\makeatother

\begin{document}
\title{Transdimensional quantum droplets in an optically trapped Bose mixture}
\author{Xiaoran Ye}
\affiliation{Department of Physics, Zhejiang Normal University, Jinhua 321004, China}
\author{Yi Zhang}
\affiliation{Department of Physics, Zhejiang Normal University, Jinhua 321004, China}
\author{Ziheng Zhou}
\affiliation{Department of Physics, Zhejiang Normal University, Jinhua 321004, China}
\author{Zhaoxin Liang}\email[Corresponding author:~] {zhxliang@zjnu.edu.cn}
\affiliation{Department of Physics, Zhejiang Normal University, Jinhua 321004, China}
\begin{abstract}
We study quantum droplets in a symmetric two-component Bose mixture with interspecies $p$-wave interactions and a two-dimensional transverse optical lattice. The lattice drives a crossover from an anisotropic three-dimensional gas to weakly coupled one-dimensional tubes. We calculate the ground-state energy and quantum depletion at the Gaussian level and derive their limiting forms. At $y=g_{12}/g=-0.95$, where the bare mean field is repulsive and no free-space droplet exists, the calculated bulk equation of state supports a self-bound minimum across the crossover: a negative lattice contribution at order $n^{2}$ supplies the attraction in the three-dimensional regime, and attractive fluctuations do so in the quasi-one-dimensional regime, with the intermediate, transdimensional range described quantitatively by neither limit. The interspecies $p$-wave interaction modifies only the spin branch. In the parameter range studied, increasing its strength lowers the equilibrium density across the crossover, consistently with a weakening of the induced binding.
\end{abstract}
\maketitle

\section{Introduction\label{sec:intro}}
Quantum droplets are self-bound states of ultracold gases. In conventional three-dimensional Bose mixtures, an attractive mean-field contribution is balanced by repulsive Lee--Huang--Yang (LHY) fluctuations~\cite{Petrov2015,Hu2020}; in the one-dimensional regime, attractive fluctuations can bind a repulsive mean-field background~\cite{Petrov2016}. They were first observed in strongly dipolar gases~\cite{FerrierBarbut2016,Schmitt2016,Chomaz2016} and subsequently in two-component Bose mixtures in both three-dimensional free space~\cite{Cabrera2018,Semeghini2018} and quasi-one-dimensional waveguides~\cite{Cheiney2018,DErrico2019}. Their equilibrium and dynamical properties, including stability, the soliton--droplet crossover, and collisions, have been extensively explored~\cite{Cabrera2018,Semeghini2018,Cheiney2018,Ferioli2019}; comprehensive reviews are given in Refs.~\cite{Bottcher2021,Chomaz2022}. Recent studies have further addressed droplet formation and fragmentation~\cite{Cavicchioli2025}, dipolar molecular droplets~\cite{Zhang2026}, liquid--gas criticality~\cite{He2023}, and droplets in nontrivial geometries~\cite{Ma2025}.

Interspecies $p$-wave interactions provide a momentum-dependent contribution to a two-component Bose mixture. Exchange symmetry forbids $p$-wave scattering between identical bosons in the same internal state, while distinguishable components can scatter in this channel. Interspecies resonances have been reported in mixtures including $^{85}\mathrm{Rb}$--$^{87}\mathrm{Rb}$ and $^{87}\mathrm{Rb}$--$^{23}\mathrm{Na}$~\cite{Dong2016,Wang2013}; the scattering description is discussed in Refs.~\cite{Idziaszek2006,Chin2010}, and $p$-wave interactions in reduced dimensions have been studied in fermionic gases~\cite{Gunter2005}. Early experiments demonstrated tunable $p$-wave interactions in ultracold Fermi gases~\cite{Regal2003}. Universal relations have been investigated experimentally in gases with $p$-wave interactions~\cite{Luciuk2016}, and $p$-wave contact relations have been derived for quasi-one-dimensional and quasi-two-dimensional traps~\cite{He2021}. For the uniform condensate considered here, the gradient vertex vanishes at mean-field level and contributes through fluctuations. Related effects on self-bound bosonic phases have been studied in Refs.~\cite{Li2019,Deng2024,Tajima2025,Ye2025}. We examine how this contribution changes as transverse tunnelling is reduced.

Transverse confinement changes the fluctuation energy and the conditions for droplet binding~\cite{Petrov2016,Ilg2018,Lavoine2021}. Bose--Bose droplets have been analysed across such a crossover under transverse harmonic confinement~\cite{Zin2018,Lavoine2021}, and in one-dimensional optical lattices in which the lattice acts along the droplet axis~\cite{Morera2020,Morera2021}; both settings involve only the $s$-wave channel.The binding mechanism differs in each case. Transverse harmonic confinement provides oscillator modes rather than a Bloch band~\cite{Zin2018,Lavoine2021,Ilg2018}, so the fluctuation sum is not restricted to a lowest band. When the lattice acts along the droplet axis, it modifies the one-dimensional equation of state directly, including a dimerized regime~\cite{Morera2021}. Here the lattice is transverse and two-dimensional: it drives the crossover itself and, through the restriction of the fluctuation sum to the lowest band, supplies a negative contribution at order $n^{2}$, so that binding becomes possible for $g_{12}>-g$. The mixture also carries an interspecies $p$-wave channel, which acts on the spin branch alone. In an optical lattice, the crossover is controlled by transverse tunnelling relative to the interaction scales~\cite{Orso2006,Hu2011,Vogler2014}. Degenerate Bose--Bose mixtures have been prepared in three-dimensional and in spin-dependent optical lattices~\cite{Catani2008,Gadway2010,Soltan-Panahi2011}, and dimensional crossovers of this type have been used to control quantum phases in dipolar gases~\cite{Biagioni2022}. Dimensional crossovers have also been investigated experimentally in ultracold Bose gases, including the 2D--1D crossover~\cite{Guo2024} and the evolution of coherence across the 1D--3D crossover~\cite{Shah2023}; a universal dimensional-crossover phase diagram has been probed with an atomic quantum simulator~\cite{Tian2026}. We use $s=2(J_1+J_2)/(gn_0)$ to label this crossover, with the tunnelling ratio specified for each scan. The density and spin branches involve the distinct scales $(1+y)gn_0$ and $(1-y)gn_0$, so their crossover ranges need not coincide. The transdimensional regime denotes the intermediate range between dimensional limits in which neither limiting description is adequate~\cite{Li2026}. Here we delineate this window quantitatively by requiring both the three-dimensional and quasi-one-dimensional asymptotic predictions for the equilibrium density to differ from the full result by more than $10\%$ at the same physical parameters.

We consider equal masses and equal intraspecies couplings, for which the density and spin modes decouple. The interspecies $p$-wave interaction modifies the spin branch. We use the resulting equation of state to follow the equilibrium density across the crossover at $y=-0.95$ and compare the full calculation with its three-dimensional and quasi-one-dimensional limits.

The paper is organized as follows. Section~\ref{sec:formalism} introduces the coherent-state path-integral formulation and derives the Bogoliubov spectrum. Section~\ref{sec:lhy} presents the beyond-mean-field corrections to the ground-state energy and quantum depletion, together with their 3D and quasi-1D asymptotic forms. Section~\ref{sec:droplets} analyzes droplet formation and stability across the dimensional crossover. Section~\ref{sec:conclusion} summarizes the main results and discusses possible directions for future work.

\section{Model and formalism\label{sec:formalism}}
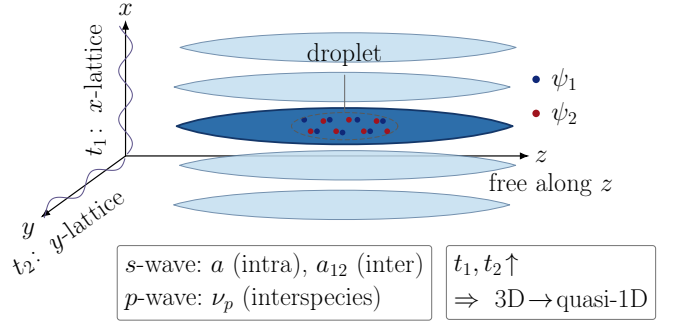
\begin{figure}[!htbp]
	\centering
	\resizebox{\columnwidth}{!}{%
		\begin{tikzpicture}[>=Latex, line cap=round, line join=round]	
			% ---- Color palette ----
			\definecolor{tubefill}{RGB}{195,222,240}
			\definecolor{tubefillhi}{RGB}{45,110,170}
			\definecolor{tubeline}{RGB}{90,125,160}
			\definecolor{tubelinehi}{RGB}{20,55,105}
			\definecolor{c1blue}{RGB}{15,40,120}
			\definecolor{c2red}{RGB}{160,20,30}
			\definecolor{latcol}{RGB}{95,85,140}
			% (0) Coordinate origin
			\coordinate (O) at (-2.5,3.4);
			% (3a) z axis
			\draw[->, line width=0.8pt] (O) -- ($(O)+(9.5,0)$)
			node[right, font=\LARGE] {$z$};
			\node[font=\LARGE, anchor=north]
			at ($(O)+(10.0,-0.30)$) {free along $z$};
			% (1) Background tubes
			\begin{scope}[opacity=0.85]
				\foreach \dy/\sx in {1.85/0.08, 0.92/-0.06, -0.92/0.07, -1.85/-0.07} {
					\draw[fill=tubefill, draw=tubeline, line width=0.6pt]
					(-1.3+\sx,4.1+\dy)
					.. controls (0.2+\sx,4.56+\dy) and (5.1+\sx,4.56+\dy) .. (6.6+\sx,4.1+\dy)
					.. controls (5.1+\sx,3.64+\dy) and (0.2+\sx,3.64+\dy) .. (-1.3+\sx,4.1+\dy)
					-- cycle;
				}
			\end{scope}
			% (2) Central tube + particles + droplet envelope
			\draw[fill=tubefillhi, draw=tubelinehi, line width=1.2pt]
			(-1.3,4.1)
			.. controls (0.2,4.67) and (5.1,4.67) .. (6.6,4.1)
			.. controls (5.1,3.53) and (0.2,3.53) .. (-1.3,4.1)
			-- cycle;
			% psi_1 (deep blue)
			\foreach \px/\py in {1.70/4.25, 2.00/3.98, 2.30/4.25, 2.60/3.95, 2.95/4.25, 3.25/3.98, 3.55/4.22}
			\fill[c1blue] (\px,\py) circle (1.9pt);
			% psi_2 (deep red)
			\foreach \px/\py in {1.85/3.98, 2.15/4.22, 2.45/3.98, 2.75/4.25, 3.10/3.98, 3.40/4.22, 3.65/3.98}
			\fill[c2red] (\px,\py) circle (1.9pt);
			% Dashed envelope
			\draw[densely dashed, draw=black!65, line width=0.6pt]
			(2.65,4.1) ellipse [x radius=1.25, y radius=0.32];
			% Droplet label
			\node[font=\LARGE, anchor=south] at (2.65,5.40) {droplet};
			\draw[line width=0.4pt, black!60] (2.65,5.30) -- (2.65,4.45);
			% Species legend
			\fill[c1blue] (7.15,5.20) circle (2.5pt);
			\node[anchor=west, font=\LARGE] at (7.40,5.20) {$\psi_1$};
			\fill[c2red]  (7.15,4.40) circle (2.5pt);
			\node[anchor=west, font=\LARGE] at (7.40,4.40) {$\psi_2$};
			% (3b) x and y axes with lattice modulations
			\draw[->, line width=0.8pt] (O) -- ($(O)+(0,3.2)$)
			node[above, font=\LARGE] {$x$};
			\begin{scope}[shift={(O)}]
				\draw[latcol, line width=0.7pt, smooth, samples=180, domain=0.05:3.05]
				plot ({0.16*sin(deg(5.5*\x))}, \x);
			\end{scope}
			\node[font=\LARGE, rotate=90, anchor=south]
			at ($(O)+(-0.35,1.55)$) {$t_1\!:\ x$-lattice};
			\draw[->, line width=0.8pt] (O) -- ($(O)+(-2.0,-1.45)$)
			node[below left, font=\LARGE] {$y$};
			\begin{scope}[shift={(O)}, rotate=215]
				\draw[latcol, line width=0.7pt, smooth, samples=160, domain=0.05:2.40]
				plot (\x, {0.16*sin(deg(5.5*\x))});
			\end{scope}
			\node[font=\LARGE, anchor=north east, rotate=35]
			at ($(O)+(-0.2,-0.55)$) {$t_2\!:\ y$-lattice};
			% (5) Crossover annotation -- bottom-right
			\node[anchor=north, font=\LARGE, align=left,
			rounded corners=2pt, draw=black!55, line width=0.4pt,
			inner sep=5pt] at (7.5,1.30)
			{$t_1,t_2\!\uparrow$\\[2pt]
				$\Rightarrow\ 3\mathrm{D}\!\to\!\mathrm{quasi}\text{-}1\mathrm{D}$};
			% (6) Interaction legend -- bottom-left
			\node[anchor=north, font=\LARGE, align=left,
			draw=black!55, rounded corners=2pt, line width=0.4pt,
			inner sep=5pt] at (1.0,1.30)
			{$s$-wave:\ $a$ (intra),\ $a_{12}$ (inter)\\
				$p$-wave:\ $\nu_p$ (interspecies)};
		\end{tikzpicture}%
	}
\caption{Schematic illustration of transdimensional quantum droplets in a two-component Bose mixture confined by a two-dimensional transverse optical lattice,
$V_{\rm opt}(x,y)=E_{\rm R}[t_1\sin^2(q_Bx)+t_2\sin^2(q_By)]$.
The lattice produces an array of quasi-one-dimensional tubes oriented along $z$, while the atoms remain free in the axial direction. Increasing the lattice depths $t_1$ and $t_2$ drives the crossover from an anisotropic three-dimensional regime to a quasi-one-dimensional regime. The two components, $\psi_1$ (blue) and $\psi_2$ (red), interact through intraspecies and interspecies $s$-wave scattering, characterized by $a$ and $a_{12}$, and through an interspecies $p$-wave interaction with scattering volume $\nu_p$. The dashed envelope indicates the self-bound droplet state considered in this work.}
	\label{fig:schematic}
\end{figure}

We consider a two-component Bose mixture subjected to a two-dimensional optical lattice in the transverse $x$--$y$ plane,
\begin{equation}
V_{\mathrm{opt}}(x,y)
=E_{\mathrm{R}}\!\left[t_{1}\sin^{2}(q_{\mathrm{B}}x)
+t_{2}\sin^{2}(q_{\mathrm{B}}y)\right],
\label{Lattice}
\end{equation}
where $t_{1}$ and $t_{2}$ are the dimensionless lattice depths,
$E_{\mathrm{R}}=\hbar^{2}q_{\mathrm{B}}^{2}/(2m)$ is the recoil energy,
$q_{\mathrm{B}}$ is the Bragg wave vector, and $m$ is the atomic mass. The lattice spacing is $d=\pi/q_{\mathrm{B}}$~\cite{Morsch2006,Bloch2008}. The atoms remain unconfined along the axial direction $z$. In the deep-lattice regime, the system therefore forms an array of weakly coupled quasi-one-dimensional tubes. Increasing $t_{1}$ and $t_{2}$ suppresses transverse motion and drives the dimensional crossover from an anisotropic three-dimensional regime to a quasi-one-dimensional regime~\cite{Orso2006,Vogler2014}. A schematic illustration is shown in Fig.~\ref{fig:schematic}.

We formulate the system within the Euclidean functional-integral framework~\cite{Yu2024,Chiquillo2018,Ye2025,Zhang2024}. The two components are represented by complex bosonic fields $\psi_i(\boldsymbol{r},\tau)$ ($i=1,2$), which depend on position $\boldsymbol{r}$ and imaginary time $\tau$ and form the spinor $\Psi=(\psi_1,\psi_2)^{T}$. The interaction comprises intraspecies and interspecies $s$-wave contact terms and an interspecies $p$-wave term. The grand-canonical partition function is
\begin{equation}
\mathcal{Z}
=\int\mathcal{D}[\bar{\Psi},\Psi]\,
\exp\!\left[-\frac{S[\bar{\Psi},\Psi]}{\hbar}\right],
\label{Parfunc}
\end{equation}
where the Euclidean action is
\begin{equation}
S[\bar{\Psi},\Psi]
=\int_{0}^{\hbar\beta}d\tau
\int_{\mathcal{V}}d\boldsymbol{r}\,
\mathcal{L}[\bar{\Psi},\Psi],
\end{equation}
and $\beta=(k_{\mathrm{B}}T)^{-1}$. The corresponding Lagrangian density reads
\begin{align}
\mathcal{L}
={}&\sum_{i=1,2}
\left[
\psi_i^{*}
\left(
\hbar\partial_{\tau}
-\frac{\hbar^{2}}{2m}\nabla^{2}
-\mu+V_{\mathrm{opt}}
\right)\psi_i
+\frac{g_{0}}{2}|\psi_i|^{4}
\right]
\nonumber\\
&+g_{012}|\psi_1|^{2}|\psi_2|^{2}
+g_{0p}
\left|
\psi_2\nabla\psi_1-\psi_1\nabla\psi_2
\right|^{2}.
\label{Lagrangian}
\end{align}
Here, $\mu$ is the chemical potential. The couplings
$g_{0}=4\pi\hbar^{2}a_{0}/m$ and
$g_{012}=4\pi\hbar^{2}a_{012}/m$
describe the intra- and interspecies $s$-wave interactions, respectively, with $a_{0}$ and $a_{012}$ the corresponding scattering lengths. In general, $g_{0}\ne g_{012}$.

We parametrize the interspecies $p$-wave interaction by
$
g_{0p}=\frac{2\pi\hbar^{2}\nu_{p}}{m},
$
where $\nu_{p}$ is the $p$-wave scattering volume~\cite{Idziaszek2006,Idziaszek2009,Deng2024}. We adopt the single-channel Idziaszek--Calarco pseudopotential, in which the three $p$-wave subchannels with $m_l=0,\pm1$ are degenerate and characterized by the same scattering volume. A molecular description of $p$-wave Feshbach resonances has also been developed~\cite{Gubbels2007}; the present calculation retains an atomic, scattering-volume-only description. The splitting induced by magnetic dipolar interactions is neglected; resolving this anisotropy within a multichannel framework is left for future work. Interspecies $p$-wave Feshbach resonances have been observed in several ultracold mixtures, including $^{85}\mathrm{Rb}$--$^{87}\mathrm{Rb}$~\cite{Dong2016} and $^{87}\mathrm{Rb}$--$^{23}\mathrm{Na}$~\cite{Wang2013}. The scattering volume can be tuned over a broad range by varying the magnetic field~\cite{Venu2023,Dong2016}. 

We work in the tight-binding regime. For transverse lattice depths $t_{1,2}\geq5$, the lowest-band approximation is valid provided that the band gap to the first excited Bloch band satisfies $E_{\mathrm{gap}}\gg\mu$~\cite{Orso2006,Li2022}. The associated Wannier orbitals are then sufficiently localized that nearest-neighbour tunnelling dominates, while phase coherence between neighbouring sites is retained~\cite{Yu2024,Hu2011}. Within the lowest band, the dimensional crossover is controlled by the ratio between the transverse bandwidth, $4(J_1+J_2)$, and the interaction scale. The mixture is anisotropically three dimensional for $\mu\ll4(J_1+J_2)$ and crosses over towards the quasi-one-dimensional regime as the transverse bandwidth is reduced relative to $\mu$.

We expand the bosonic fields in orthonormal Wannier orbitals $W_j$ of the lowest Bloch band, normalized by $\int_{\mathbb R}dx_j\,|W_j(x_j)|^2=1$. For the on-site interaction integrals, we approximate $W_j$ by the normalized Gaussian $\omega_j$. The Bloch expansion and the Gaussian approximation are
\begin{align}
\phi_{k_j}(x_j)
&=\frac{1}{\sqrt{N_j}}\sum_l
e^{i l d k_j}\,W_j(x_j-ld),\\
\omega_j(x_j)
&=\frac{1}{\pi^{1/4}\sigma_j^{1/2}}
\exp\!\left(-\frac{x_j^2}{2\sigma_j^2}\right),
\end{align}
Here $j=1,2$, $x_1=x$, $x_2=y$, and $N_j=L_j/d$ is the number of sites in direction $j$. The Gaussian widths are parameterized by
\begin{equation}
\frac{d}{\sigma_j}
=\pi t_j^{1/4}
\exp\!\left(-\frac{1}{4\sqrt{t_j}}\right).
\end{equation}
The bosonic field is expanded as
\begin{equation}
\psi_i(\boldsymbol{r},\tau)
=\frac{1}{\sqrt{L_z}}
\sum_{\boldsymbol{k},n}
a_{i\boldsymbol{k},n}\,
\phi_{k_x}(x)\phi_{k_y}(y)
e^{-ik_z z+i\omega_n\tau},
\end{equation}
where $\omega_n=2\pi n/(\hbar\beta)$ are bosonic Matsubara frequencies. The action of the model system then takes the form
\begin{widetext}
\begin{align}
\frac{S}{\hbar\beta}
={}&\sum_{i,\boldsymbol{k},n}
a_{i\boldsymbol{k},n}^{*}
\left(-i\hbar\omega_n+\varepsilon_{\boldsymbol{k}}-\mu\right)
a_{i\boldsymbol{k},n}
\nonumber\\
&+\frac{g}{2V}
\sum_i
\sum_{\substack{\boldsymbol{k},\boldsymbol{k}',\boldsymbol{q}\\ n,n',m}}
a_{i,\boldsymbol{k}+\boldsymbol{q},n+m}^{*}
a_{i,\boldsymbol{k}'-\boldsymbol{q},n'-m}^{*}
a_{i,\boldsymbol{k},n'}
a_{i,\boldsymbol{k}',n}+\frac{g_{12}}{V}
\sum_{\substack{\boldsymbol{k},\boldsymbol{k}',\boldsymbol{q}\\ n,n',m}}
a_{1,\boldsymbol{k}+\boldsymbol{q},n+m}^{*}
a_{2,\boldsymbol{k}'-\boldsymbol{q},n'-m}^{*}
a_{1,\boldsymbol{k},n'}
a_{2,\boldsymbol{k}',n}
\nonumber\\
&+\frac{g_p}{V}
\sum_{\substack{\boldsymbol{k},\boldsymbol{k}',\boldsymbol{q}\\ n,n',m}}
(k'_z-k_z-2q_z)(k'_z-k_z)
a_{2,\boldsymbol{k}+\boldsymbol{q},n+m}^{*}
a_{1,\boldsymbol{k}'-\boldsymbol{q},n'-m}^{*}
a_{2,\boldsymbol{k},n'}
a_{1,\boldsymbol{k}',n}.
\label{Action}
\end{align}
\end{widetext}

The single-particle dispersion is
\begin{equation}
\varepsilon_{\boldsymbol{k}}
=\frac{\hbar^2 k_z^2}{2m}
+2\left[
J-J_1\cos(k_xd)-J_2\cos(k_yd)
\right],
\end{equation}
where $J=J_1+J_2$. The transverse nearest-neighbour tunnelling amplitudes are the matrix elements between orthonormal Wannier orbitals,
\begin{equation}
\begin{aligned}
J_j={}&-\int_{-\infty}^{\infty}dx_j\,
W_j^*(x_j)
\left[-\frac{\hbar^2}{2m}\partial_{x_j}^2\right.\\
&\left.\hspace{3.0em}{}+E_{\mathrm R}t_j\sin^2(q_{\mathrm B}x_j)\right]
W_j(x_j-d).
\end{aligned}\label{Jdef}
\end{equation}
For conversions between lattice depth and tunnelling, we use the deep-lattice band asymptote $J_j/E_R\simeq4t_j^{3/4}e^{-2\sqrt{t_j}}/\sqrt\pi$~\cite{Bloch2008,Li2022}. At $t_j=5$, this gives $J_j/E_R\simeq0.086$. The Gaussian approximation is used for the on-site interaction integrals; its intersite matrix elements are not used to estimate $J_j$.
Evaluating the on-site interaction integrals with the Gaussian orbitals gives the effective couplings
\begin{equation}
g=\frac{g_0d^2}{2\pi\sigma_x\sigma_y},
\qquad
g_{12}=\frac{g_{012}d^2}{2\pi\sigma_x\sigma_y},
\qquad
g_p=\frac{g_{0p}d^2}{2\pi\sigma_x\sigma_y}.\label{gproj}
\end{equation}
The Gaussian normalization gives $\int_{\mathbb R}dx_j\,|\omega_j|^4=1/(\sqrt{2\pi}\sigma_j)$. Thus the coupling within one tube is $g^{\mathrm{1D}}=g_0/(2\pi\sigma_x\sigma_y)$. With $V=N_1N_2d^2L_z$, the convention $g/(2V)$ in Eq.~(\ref{Action}) gives $g=d^2g^{\mathrm{1D}}$. The factor $d^2$ therefore converts the tube coupling to the three-dimensional volume convention.
The transverse components of the projected $p$-wave vertex cancel because of its antisymmetric structure and the identical transverse Wannier orbitals of the two components. Only the axial component survives. We further assume a sufficiently large occupation per tube that the Mott-insulator transition in the transverse lattice can be neglected~\cite{Oosten2003,Greiner2002}.

We restrict the analysis to a symmetric mixture with equal masses,
$m_1=m_2\equiv m$, equal intraspecies scattering lengths,
$a_1=a_2\equiv a_{0}$, balanced populations,
$N_{01}=N_{02}\equiv N_0$, and equal chemical potentials,
$\mu_1=\mu_2\equiv\mu$. Under these conditions, the density and spin modes decouple, and the interspecies $p$-wave interaction enters only the spin sector at Gaussian order.

Quantum fluctuations are included within the one-loop approximation. We expand the lattice-projected fields about a uniform condensate,
\begin{equation}
a_{i\boldsymbol{k},n}
=\sqrt{N_0}\,\delta_{\boldsymbol{k},0}\delta_{n,0}
+\varphi_{i\boldsymbol{k},n},
\end{equation}
where $\varphi_{i\boldsymbol{k},n}$ describes fluctuations about the zero-momentum, zero-frequency condensate mode. The condensate density per component is $n_0=N_0/L^3$, with $L^3=L_xL_yL_z$. In the deep-lattice limit, it is related to the line density in each tube by
\begin{equation}
n_{0\mathrm{1D}}=n_0d^2.
\end{equation}
Retaining terms up to second order in the fluctuation fields gives
\begin{equation}
S=S_0+S_g+\cdots,
\end{equation}
where $S_0$ is the mean-field action and $S_g$ contains the Gaussian fluctuations. Stationarity of $S_0$ yields
\begin{equation}
\mu=(g+g_{12})n_0.
\end{equation}
The $p$-wave interaction does not contribute to this saddle-point equation because its gradient vertex vanishes for a uniform condensate. It therefore enters the energy only through the Gaussian sector.

The Gaussian action can be written as
\begin{equation}
\frac{S_g}{\hbar\beta}
=\frac{1}{2}\sum_{\boldsymbol{k},n}
\boldsymbol{\Phi}_{\boldsymbol{k}n}^{\dagger}
\boldsymbol{M}'(i\hbar\omega_n,\boldsymbol{k})
\boldsymbol{\Phi}_{\boldsymbol{k}n},
\end{equation}
where
\begin{equation}
\boldsymbol{M}'(i\hbar\omega_n,\boldsymbol{k})
=-i\hbar\omega_n\boldsymbol{\kappa}
+\boldsymbol{M}(\boldsymbol{k}),
\qquad
\boldsymbol{\kappa}
=\begin{pmatrix}
\sigma_z & 0\\
0 & \sigma_z
\end{pmatrix},
\end{equation}
and
\begin{equation}
\boldsymbol{\Phi}_{\boldsymbol{k}n}
=
\begin{pmatrix}
\varphi_{1\boldsymbol{k},n}\\
\varphi_{1,-\boldsymbol{k},-n}^{*}\\
\varphi_{2\boldsymbol{k},n}\\
\varphi_{2,-\boldsymbol{k},-n}^{*}
\end{pmatrix}
\end{equation}
is the Nambu spinor. Defining
\begin{align}
A_{\boldsymbol{k}}
&=\varepsilon_{\boldsymbol{k}}-\mu
+2gn_0+g_{12}n_0+g_pn_0k_z^2,\\
B&=gn_0,\\
C_{\boldsymbol{k}}
&=g_{12}n_0-g_pn_0k_z^2,\\
D&=g_{12}n_0,
\end{align}
the Bogoliubov--de Gennes kernel is
\begin{equation}
\boldsymbol{M}(\boldsymbol{k})
=
\begin{pmatrix}
A_{\boldsymbol{k}} & B & C_{\boldsymbol{k}} & D\\
B & A_{\boldsymbol{k}} & D & C_{\boldsymbol{k}}\\
C_{\boldsymbol{k}} & D & A_{\boldsymbol{k}} & B\\
D & C_{\boldsymbol{k}} & B & A_{\boldsymbol{k}}
\end{pmatrix}.
\label{Mk}
\end{equation}

Setting $g_p=0$ in Eq.~(\ref{Mk}) gives the $s$-wave mixture with the lattice dispersion retained~\cite{Cappellaro2017,Chiquillo2018}. At $g_{12}=g_p=0$, the two components decouple into independent Bose gases in the same lattice. The isotropic free-space theory~\cite{Ye2025} must be obtained from the unprojected Lagrangian in Eq.~(\ref{Lagrangian}) by setting $V_{\mathrm{opt}}=0$ and retaining all three spatial components of the $p$-wave vertex. This limit cannot be taken directly in the axial-vertex, lowest-band kernel of Eq.~(\ref{Mk}).

The elementary Bogoliubov modes follow from the static BdG kernel $\boldsymbol{M}(\boldsymbol{k})$. The Cayley--Hamilton theorem~\cite{Wang2022} gives the upper ($\mathrm{u}$) and lower ($\mathrm{l}$) energies as
\begin{equation}
\begin{aligned}
E_{\mathrm{u}/\mathrm{l}}
&=\Bigg\{\frac{1}{4}\operatorname{Tr}\!\left[(\boldsymbol{\kappa M})^{2}\right] \\
&\quad\pm\sqrt{\frac{1}{16}\left[\operatorname{Tr}\!\left((\boldsymbol{\kappa M})^{2}\right)\right]^{2}-\det(\boldsymbol{\kappa M})}\Bigg\}^{1/2}.
\end{aligned}
\end{equation}

We label the density (sum) mode by $+$ and the spin (density-difference) mode by $-$. For $-1<y<0$ and $z\ge0$, their energies satisfy $E_{-}=E_{\mathrm{u}}\ge E_{\mathrm{l}}=E_{+}$. In dimensionless form,
\begin{equation}
E_{\pm} =gn_{0}f_{\pm},
\end{equation}
where $f_{\pm}$ denote the two branches of the dimensionless excitation spectrum. The density branch is
\begin{equation*}
f_{+}=\sqrt{\left(k_z^{\prime2}+s\gamma\right)\left(2+k_z^{\prime2}+s\gamma+2y\right)},
\end{equation*}
while the spin branch is
\begin{equation*}
\begin{aligned}
f_{-}&=\Big\{\left[k_z^{\prime2}(1+2z)+s\gamma\right]\\
&\quad\times\left[k_z^{\prime2}(1+2z)+s\gamma+2(1-y)\right]\Big\}^{1/2}.
\end{aligned}
\end{equation*}

Here $k_z^{\prime}=\hbar k_z/\sqrt{2mgn_0}$ is the dimensionless axial momentum, $s=2J/(gn_{0})$, $\gamma=1-\frac{J_{1}}{J}\cos(k_{x}d)-\frac{J_{2}}{J}\cos(k_{y}d)$, $y=g_{12}/g$, and $z=2mg_{p}n_{0}/\hbar^{2}$. The dimensionless $p$-wave parameter $z$ enters only the spin branch $f_{-}$ and is absent from the density branch $f_{+}$. This reflects the antisymmetric structure of the interspecies $p$-wave vertex in the species index: it dresses the spin (density-difference) mode but leaves the density (sum) mode untouched, mirroring the absence of $g_{p}$ from the saddle-point relation.

Writing $X=k_z^{\prime2}+s\gamma$ and $\widetilde X=k_z^{\prime2}(1+2z)+s\gamma$, the two branches obey $f_+^2=X[X+2(1+y)]$ and $f_-^2=\widetilde X[\widetilde X+2(1-y)]$. For $y<1$ and a positive axial kinetic coefficient, both branches are real at all momenta under the conditions
\begin{equation}
1+y\ge0\qquad\text{and}\qquad1+2z>0 .\label{reality}
\end{equation}
The calculations below use $-1<y<0$ and $z\ge0$. The boundary $y=-1$ is independent of the lattice depths.

\section{Lee-Huang-Yang corrections and quantum depletion\label{sec:lhy}}

We now calculate the zero-temperature fluctuation energy and quantum depletion from the Bogoliubov spectrum within the Gaussian approximation~\cite{Lee1957,Beliaev1958,Andersen2004,Salasnich2016}.

Our starting point is the Gaussian-fluctuation grand potential $\Omega_{g}\left(\mu,n_{0}\right)$, which follows from the inverse Green’s function $\boldsymbol{M}^{\prime}\left(k,i\omega_{n}\right)$ through the standard trace-log expression and the subsequent summation
over Matsubara frequencies. It can be written as
\begin{equation}
\Omega_{g}\left(\mu,n_{0}\right)  =\frac{1}{2}\sum_{\boldsymbol k,\sigma}\left[E_{\boldsymbol k,\sigma}-\mathcal{A}_{\boldsymbol k,\sigma}+\frac{2}{\beta}\ln\left(1-e^{-\beta E_{\boldsymbol k,\sigma}}\right)\right],\label{Omegag}
\end{equation}

Here $\sigma=\pm1$ labels the density and spin branches, and the condensate mode is excluded from the fluctuation sums. The Gaussian determinant is evaluated with the time-ordering prescription inherited from the discretized coherent-state path integral of the normal-ordered Hamiltonian~\cite{Salasnich2016}. This fixes the vacuum contribution to $\frac12\sum_{\boldsymbol k,\sigma}(E_{\boldsymbol k,\sigma}-\mathcal A_{\boldsymbol k,\sigma})$, where $\mathcal A_{\boldsymbol k,\sigma}=A_{\boldsymbol k}+\sigma C_{\boldsymbol k}$. The subtraction is distinct from coupling-constant renormalization and is retained in the thermodynamic derivatives below; its $s$-wave lattice counterpart appears in Ref.~\cite{Orso2006}.

At zero temperature, we evaluate the energy through the Legendre transform $E_g=\Omega_0+\Omega_g^{(0)}+2L^3\mu n_0$. To the order retained here, the energy density is
\begin{align}
\frac{E_{g}}{L^{3}} & =gn^{2}_{0}+g_{12}n^{2}_{0}-\frac{\left(gn_{0}\right)^{3/2}}{4\pi d^{2}}\left(\frac{2m}{\hbar^{2}}\right)^{1/2}f(s),\label{Eg}
\end{align}
The first two terms are the mean-field energy. The fluctuation contribution is expressed through the dimensionless function
\begin{widetext}
\begin{align}
f(s)={}&\int_{-\pi}^{\pi} d^2 k\,\Bigg\{
\frac{\sqrt{s\gamma}}{6\pi}(1+y+s\gamma)
{}_2F_1\!\left(-\frac12,\frac12;1;-\frac{2(1+y)}{s\gamma}\right)
-\frac{\sqrt{s\gamma}}{6\pi}(2+2y+s\gamma)
{}_2F_1\!\left(\frac12,\frac12;1;-\frac{2(1+y)}{s\gamma}\right)
\nonumber\\
&+\frac{\sqrt{s\gamma}}{6\pi\sqrt{1+2z}}(1-y+s\gamma)
{}_2F_1\!\left(-\frac12,\frac12;1;\frac{2(y-1)}{s\gamma}\right)-\frac{\sqrt{s\gamma}}{6\pi\sqrt{1+2z}}(2-2y+s\gamma)
{}_2F_1\!\left(\frac12,\frac12;1;\frac{2(y-1)}{s\gamma}\right)
\Bigg\}.
\label{fs}
\end{align}
\end{widetext}

Here $_{2}F_{1}$ denotes the Gauss hypergeometric function, and the derivation of Eq.~(\ref{fs}) is presented in Appendix~\ref{A}. The four terms in Eq.~(\ref{fs}) correspond to the density branch $E_{+}$ (first two terms) and the spin branch $E_{-}$ (last two terms). The $p$-wave parameter $z$ enters only through the prefactor $1/\sqrt{1+2z}$ of the spin-branch terms, in agreement with Sec.~\ref{sec:formalism}. 

It is useful to record the large-momentum behaviour of the normal-ordered integrand. Writing $\mathcal A_{\boldsymbol k,\sigma}=X_{\boldsymbol k,\sigma}+\lambda_{\sigma}$ with $X_{\boldsymbol k,+}=\varepsilon_{\boldsymbol k}$, $X_{\boldsymbol k,-}=\varepsilon_{\boldsymbol k}+2g_{p}n_{0}k_{z}^{2}$ and $\lambda_{\pm}=(g\pm g_{12})n_{0}$, the Bogoliubov energies are $E_{\boldsymbol k,\sigma}=[X_{\boldsymbol k,\sigma}(X_{\boldsymbol k,\sigma}+2\lambda_{\sigma})]^{1/2}$ and $\mathcal A_{\boldsymbol k,\sigma}-E_{\boldsymbol k,\sigma}=\lambda_{\sigma}^{2}/(2X_{\boldsymbol k,\sigma})-\lambda_{\sigma}^{3}/(2X_{\boldsymbol k,\sigma}^{2})+\cdots$, so that the integrand decays as $k_{z}^{-2}$ at large axial momentum for the parameters considered here. The transverse quasimomenta are restricted to the first Brillouin zone, so the lowest-band fluctuation integral is ultraviolet convergent and we evaluate it without any further subtraction~\cite{Salasnich2016}. In particular, the finite second-order contribution $\lambda_{\sigma}^{2}/(2X_{\boldsymbol k,\sigma})$ is retained; it is the origin of the term of order $n^{2}$ in Eq.~(\ref{Eg3D}) and, in the $s$-wave limit, corresponds to the lattice correction identified in Ref.~\cite{Orso2006}. By contrast, in the free-space calculation of Ref.~\cite{Ye2025} the same contribution is ultraviolet divergent and is absorbed into the renormalization of the coupling constant~\cite{Braaten1997}; that step is not repeated within the band. The $p$-wave channel is treated in the scattering-volume-only approximation throughout. The term $2g_{p}n_{0}k_{z}^{2}$ in $X_{\boldsymbol k,-}$ improves the convergence of the spin-branch integrand, but this is separate from the validity of that approximation, which is controlled by the effective range in the same large-$k_z$ region. Performing the $k_{z}$ integration analytically then yields the hypergeometric scaling function $f(s)$ in Eq.~(\ref{fs}).

The physical input of the lowest-band description consists of the renormalized free-space couplings $g_0=4\pi\hbar^2a_0/m$ and $g_{012}=4\pi\hbar^2a_{012}/m$, whose Wannier projection gives Eq.~(\ref{gproj}) and the effective lengths defined after Eq.~(\ref{Eg3D}); the projection is used for $a_0\ll\sigma_j$, and the three-dimensional and crossover plots use fixed projected lengths. For $J_1,J_2>0$, the lowest-band integral of $1/(2\varepsilon_{\boldsymbol k})$ is finite. Within this convention the finite second-order contribution is kept explicitly in the loop correction; at a given order it could equivalently be redistributed into a redefined low-energy coupling, provided the same convention is used throughout, so that no contribution is counted twice. The finite lattice contribution retained in Eq.~(\ref{Eg3D}) enters at order $n^2$. For $J_1=J_2$, its geometric coefficient is
\begin{equation}
\begin{aligned}
C_{\mathrm{lat}}&=\frac{\pi}{2(2\pi)^2}\int_{-\pi}^{\pi}
\frac{d^2q}{\sqrt{2-\cos q_x-\cos q_y}}\\
&\simeq1.428,
\end{aligned}\label{lattconst}
\end{equation}
where $\boldsymbol q=(k_xd,k_yd)$ is dimensionless. We use the rounded value $1.43$ below, consistent with Ref.~\cite{Orso2006}. The corresponding integral depends on the tunnelling ratio in an anisotropic lattice.

In the limit $y=0$ and $z=0$, the ground-state energy of Eq.~(\ref{Eg}) reduces to
\begin{equation}
\frac{E^{\left(0\right)}_{\text{g}}}{L^{3}}  =gn^{2}_{0}-\frac{\left(gn_{0}\right)^{\frac{3}{2}}}{4\pi d^{2}}\left(\frac{2m}{\hbar^{2}}\right)^{\frac{1}{2}}f^{\left(0\right)}(s),\label{Eg0}
\end{equation}
where $f^{(0)}(s)$ is
\begin{eqnarray}
f^{\left(0\right)}\left(s\right) &=&\int^{\pi}_{-\pi}d^{2}k\Bigg\{\frac{\sqrt{s\gamma}}{3\pi}\left(1+s\gamma\right){}_{2}F_{1}\left[-\frac{1}{2},\frac{1}{2},1,\frac{-2}{s\gamma}\right]\nonumber \\
&-&\frac{\sqrt{s\gamma}}{3\pi}\left(2+s\gamma\right){}_{2}F_{1}\left[\frac{1}{2},\frac{1}{2},1,\frac{-2}{s\gamma}\right]\Bigg\}.\label{fs0}
\end{eqnarray}

At $y=z=0$, Eq.~(\ref{fs0}) recovers the lattice result of Ref.~\cite{Li2022}. Equation~(\ref{Eg}) applies within the lowest-band projection. The isotropic free-space energy requires a separate calculation from Eq.~(\ref{Lagrangian}) with $V_{\mathrm{opt}}=0$~\cite{Ye2025}.

Quantum depletion characterizes the fraction of particles occupying excited states outside the condensate. At zero temperature, it can be obtained from the grand potential through the thermodynamic identity $N=-\partial\left(\Omega_{0}+\Omega_{g}\right)/\partial\mu$, where $\Omega_{0}$ denotes the mean-field contribution and $\Omega_{g}$ represents the Gaussian fluctuation correction. The derivative is taken at fixed $n_0$, $g$, $g_{12}$, and $g_p$, before imposing $\mu=(g+g_{12})n_0$. The normal-ordering term of Eq.~(\ref{Omegag}) contains no fluctuation operators but depends on $\mu$, and is therefore differentiated together with $E_{\boldsymbol k,\sigma}$. Since $\partial_\mu\mathcal A_{\boldsymbol k,\sigma}=-1$ and $\partial_\mu E_\sigma=-\mathcal A_{\boldsymbol k,\sigma}/E_\sigma$, the total density is
\begin{equation*}
\begin{aligned}
n&=2n_0-\frac{1}{2V}\sum_{\boldsymbol k,\sigma}
\partial_\mu\big(E_\sigma-\mathcal A_{\boldsymbol k,\sigma}\big)\\
&=2n_0+\frac{1}{2V}\sum_{\boldsymbol k,\sigma}
\left(\frac{\mathcal A_{\boldsymbol k,\sigma}}{E_\sigma}-1\right).
\end{aligned}
\end{equation*}
The last expression is evaluated at the mean-field saddle point. The $-1$ from each branch gives the $-2$ in Eq.~(\ref{hs}); the prefactor of Eq.~(\ref{n}) uses $n\simeq2n_0$ to leading order in the depletion. By converting the momentum summation into an integral within the path-integral formalism, we finally obtain an analytical expression for the quantum depletion, 
\begin{equation}
\frac{N-2N_{0}}{N}  =\frac{1}{4\pi d^{2}}\left(\frac{mg}{\hbar^{2}n}\right)^{\frac{1}{2}}h(s),\label{n}
\end{equation}
where $h\left(s\right)$ is given by
\begin{widetext}
\begin{equation}
h\left(s\right) =\frac{1}{2\pi^{2}}\int d^{3}\mathbf{k}\Bigg\{\frac{k^{\prime2}_{z}+s\gamma+1+y}{\sqrt{\left(k^{\prime2}_{z}+s\gamma\right)\left[k^{\prime2}_{z}+s\gamma+2\left(1+y\right)\right]}}
\!\!+  \!\!\frac{k^{\prime2}_{z}\left(1+2z\right)+s\gamma+1-y}{\sqrt{\left(k^{\prime2}_{z}\left(1+2z\right)+s\gamma\right)\left(k^{\prime2}_{z}\left(1+2z\right)+s\gamma+2\left(1-y\right)\right)}}-2\Bigg\},\label{hs}
\end{equation}
\end{widetext}
with $\int d^{3}\mathbf{k}=\int^{\pi}_{-\pi}d^{2}k\int^{\infty}_{0}dk^{\prime}_{z}$, $k^{\prime}_{z}=\hbar k_{z}/(\sqrt{2mgn_{0}})$ is the dimensionless axial momentum. By setting $y=0$ and $z=0$, Eq.~(\ref{n}) simplifies to the form:
\begin{equation}
\frac{N-2N_{0}}{N}  =\frac{1}{4\pi d^{2}}\left(\frac{mg}{\hbar^{2}n}\right)^{\frac{1}{2}}h^{\left(0\right)}(s),\label{n0}
\end{equation}
where $h^{\left(0\right)}\left(s\right)$ reads 
\begin{equation}
h^{\left(0\right)}\left(s\right)  =\frac{1}{\pi^{2}}\int d^{3}\mathbf{k}\left[\frac{k^{\prime2}_{z}+s\gamma+1}{\sqrt{\left(k^{\prime2}_{z}+s\gamma\right)\left[k^{\prime2}_{z}+s\gamma+2\right]}}-1\right].\label{hs0}
\end{equation}

\begin{figure}[t] 
 \begin{centering} 
  \includegraphics[scale=0.46]{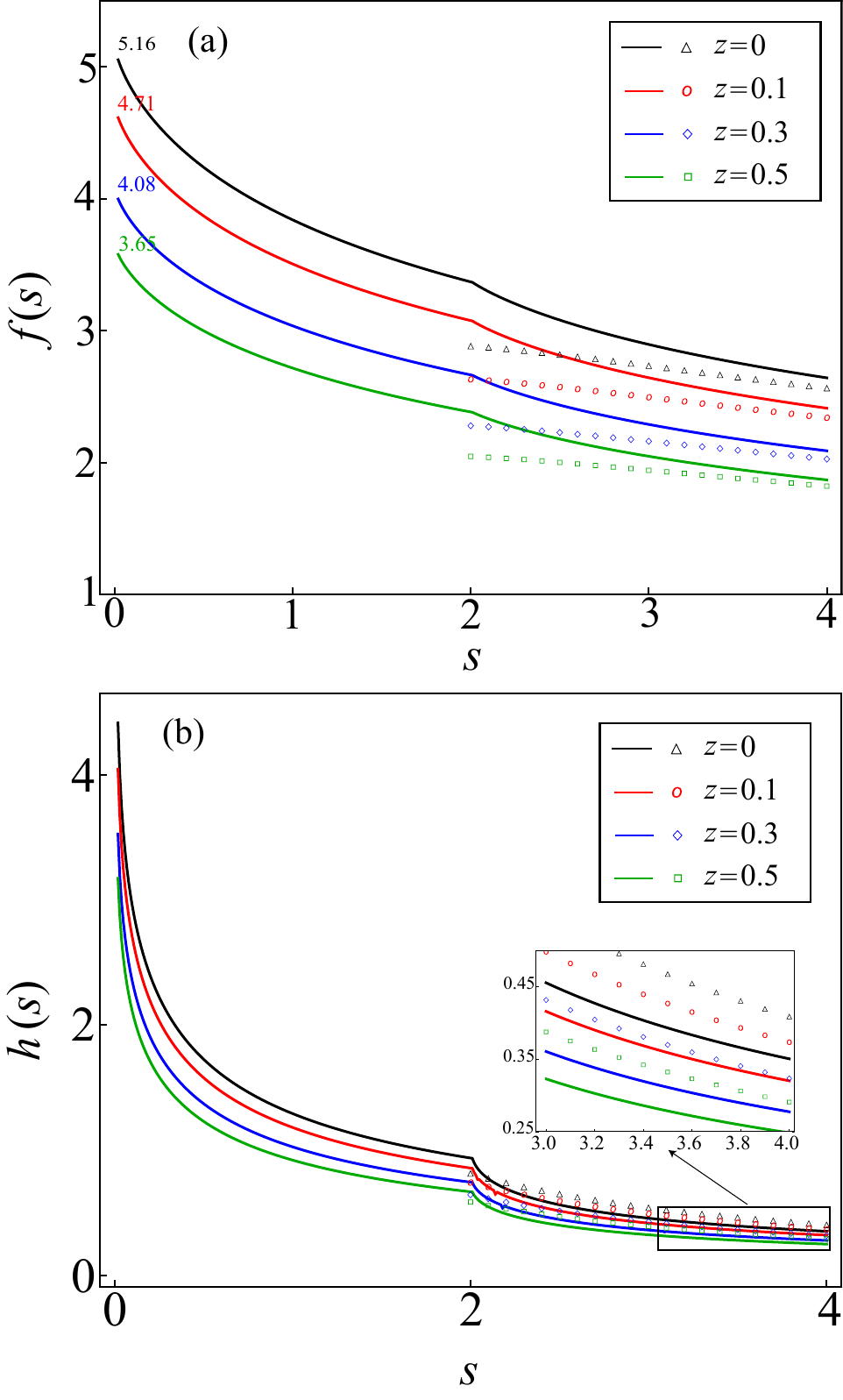} 
  \end{centering} 
 \caption{(a) Scaling function $f(s)$ and (b) scaling function $h(s)$ for different values of the dimensionless $p$-wave interaction strength $z$, computed at $y=-0.95$. The dimensional crossover is traversed in two stages: first, $s_2=0.01$ is held fixed while $s_1$ is varied from 0 to 2, driving the crossover from the quasi-1D regime to a quasi-2D regime; then $s_1=2$ is held fixed while $s_2$ is varied from 0.01 to 2, further driving the crossover from the quasi-2D regime to the anisotropic 3D regime. Solid curves show the full results from Eqs.~(\ref{fs}) and~(\ref{hs}); the symbols on the right mark the 3D asymptotic values, and the numbers on the left indicate the quasi-1D limiting values. Increasing $z$ suppresses both $f(s)$ and $h(s)$. The suppression enters through the spin-branch terms alone and is therefore not uniform in $s$.
    \label{fig:fshs}
}
\end{figure}

The $y=z=0$ expressions recover Ref.~\cite{Li2022}. Figure~\ref{fig:fshs} shows $f$ and $h$ along the tunnelling path specified in the caption. The dependence on $z$ enters through the spin-branch factor $(1+2z)^{-1/2}$. The energy function has a finite one-dimensional limit. At fixed $y$ and $z$, the depletion function instead diverges logarithmically when both transverse tunnelling amplitudes vanish at a fixed nonzero ratio: $h(s)=-A(y,z)\ln s+O(1)$, with $A(y,z)=\sqrt{(1+y)/2}+\sqrt{(1-y)/2}/\sqrt{1+2z}$. The depletion at finite tunnelling must therefore be evaluated at the densities used in the droplet calculation.

Negative $y=g_{12}/g$ describes attractive interspecies and repulsive intraspecies interactions. The conventional free-space three-dimensional droplet has $y<-1$~\cite{Petrov2015}, whereas the lattice equation of state considered below supports binding at $y=-0.95$. Feshbach resonances in $^{39}\mathrm{K}$ have been characterized spectroscopically~\cite{DErrico2007}. Feshbach tuning can vary the scattering parameters~\cite{Chin2010,Roati2007,Thalhammer2008}, and the $s$-wave control assumed here has been characterized in $^{39}\mathrm{K}$ Bose--Bose mixtures~\cite{Tanzi2018} and in heteronuclear mixtures~\cite{Ospelkaus2006}, but identifying a particular mixture requires compatible $s$- and $p$-wave parameters. The dimensionless quantity $z=2mg_pn_0/\hbar^2$ also varies with density at fixed $g_p$.

\section{Quantum droplets in the 3D--1D dimensional crossover\label{sec:droplets}}

We next derive the limiting equations of state and identify their self-bound minima. The parameters $s_j=2J_j/(gn_0)$ compare tunnelling in each transverse direction with the interaction scale~\cite{Cazalilla2006,Bloch2008,Li2022,Yu2024}. The sum $s=s_1+s_2$ labels a scan with a specified tunnelling ratio or path. The lowest-band description requires the band gap to exceed the interaction energies, as discussed in Sec.~\ref{sec:formalism}.

\subsection{3D asymptotic behavior\label{sec:3D}}

For the symmetric transverse lattice, $J_1=J_2$, and sufficiently large $s$, the scaling function has the three-dimensional asymptotic expansion
\begin{align}
	f_{\text{3D}}\left(s\right) & =-\frac{32\sqrt{2}}{15\pi s}\left(\left(y+1\right)^{5/2}+\frac{(1-y)^{5/2}}{\sqrt{1+2z}}\right)\nonumber \\
	& +\frac{1.43\sqrt{2}}{\sqrt{s}}\left(\left(1+y\right)^{2}+\frac{\left(1-y\right)^{2}}{\sqrt{1+2z}}\right).\label{fs3D}
\end{align}

\begin{figure}[htbp] 
 \begin{centering} 
  \includegraphics[scale=0.46]{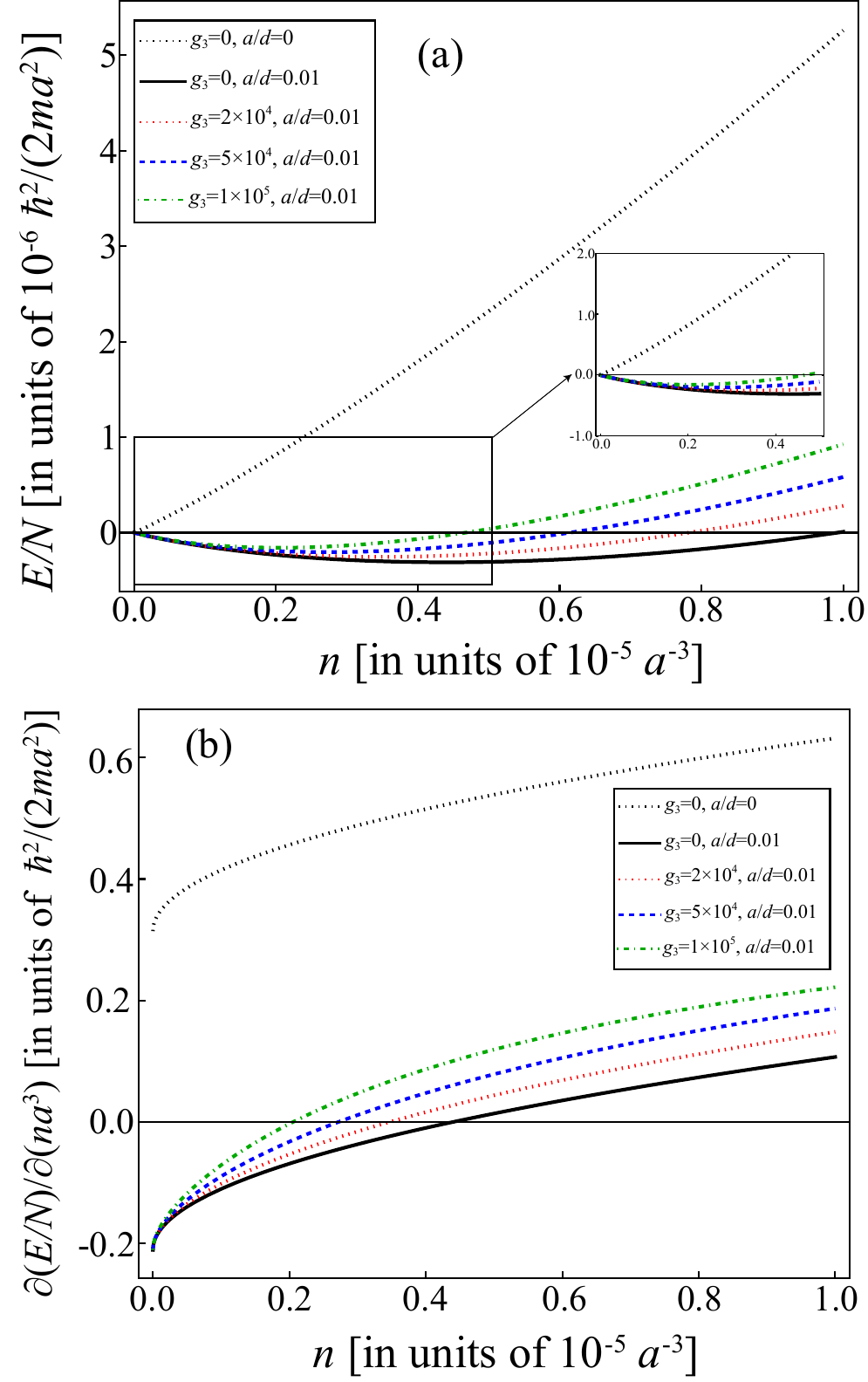} 
  \end{centering} 
 \caption{(a) Asymptotic 3D energy per particle $E/N$ (in units of $10^{-6}\hbar^{2}/2ma^{2}$) as a function of the density $n$ (in units of $10^{-5}a^{-3}$) for $g_{3}=0$ (black solid), $2\times10^{4}$ (red dotted), $5\times10^{4}$ (blue dashed), and $10^{5}$ (green dot-dashed), at $y=-0.95$, $a/d=0.01$, and $m^{*}/m\approx1.18$ ($t_{1,2}=5$). A finite-density minimum signals the self-bound droplet state; the inset enlarges the low-density region. (b) Corresponding derivative $\partial(E/N)/\partial(na^{3})$ (in units of $\hbar^{2}/2ma^{2}$), whose zero crossing determines the equilibrium droplet density. As $g_{3}$ increases, the minimum in panel (a) and the zero crossing in panel (b) shift toward lower density and the binding becomes shallower.The curve labelled $a/d=0$, shown in both panels, is a formal reference obtained by omitting the lattice-induced $n^2$ term from the asymptotic equation of state at $g_3=0$. It has no self-bound minimum at $y=-0.95$, and its derivative in panel (b) has no zero crossing. This comparison isolates that term within the lattice model; it does not take the physical limit $V_{\mathrm{opt}}\to0$.
    \label{fig:3Ddroplet}
}
\end{figure}

The two terms in Eq.~(\ref{fs3D}) have different physical origins. The first term, proportional to $s^{-1}$, is the generalized three-dimensional LHY correction. The second, proportional to $s^{-1/2}$, is a lattice-induced renormalization of the interaction strength~\cite{Orso2006}, modified by the interspecies $p$-wave coupling. This contribution is itself a zero-point fluctuation effect. The lowest-band Brillouin-zone sum shifts part of the fluctuation energy to the same order in density as the bare mean-field term and gives a negative contribution, thereby producing an effective attraction even when the bare mean field, proportional to $1+y$, is repulsive. The usual LHY term enters at higher order in density and provides the stabilizing repulsion.
The coefficient in Eq.~(\ref{fs3D}) is defined in Eq.~(\ref{lattconst}). It is the symmetric-lattice value, and Eqs.~(\ref{fs3D})--(\ref{n3D}) below, together with $m^{*}=\hbar^{2}/(Jd^{2})$, are written for $J_1=J_2$; the scans of Figs.~\ref{fig:3Ddroplet}--\ref{fig:crossover} are symmetric. For an anisotropic lattice the integral of Eq.~(\ref{lattconst}) depends on the tunnelling ratio, and the general two-parameter path of Fig.~\ref{fig:fshs} is evaluated from the full $f(s)$ rather than from this expansion.

The large-$s$ limit remains an anisotropic three-dimensional lattice gas. Within the on-site projection onto identical transverse Wannier orbitals, the $p$-wave vertex has only an axial component; its spin-branch contribution therefore carries the factor $(1+2z)^{-1/2}$. Removing the lattice from the unprojected model restores the transverse components of the vertex and leads to the isotropic free-space result~\cite{Ye2025}. Increasing $s$ within the lowest band does not perform this removal.

Substituting Eq.~(\ref{fs3D}) into the general expression for the ground-state energy density Eq.~(\ref{Eg}), we obtain the asymptotic energy density in the 3D regime as:
\begin{eqnarray}
\!\!\frac{E^{\text{3D}}_{\text{g}}}{L^{3}}  &\!=\!&\frac{4\pi\hbar^{2}a}{m}n^{2}_{0}\Bigg\{1+y-\frac{a}{a_{\text{cr}}}\nonumber \\
&\!+\!&\frac{64}{15}\frac{m^{*}}{m}\sqrt{\frac{n_{0}a^{3}}{\pi}}\left[\left(1+y\right)^{\frac{5}{2}}+\frac{(1-y)^{\frac{5}{2}}}{\left(1+2z\right)^{\frac{1}{2}}}\right]\Bigg\},
\label{Eg3D}
\end{eqnarray}
with $m^{*}=\hbar^{2}/(Jd^{2})$, $a_{\text{cr}}=\sqrt{m}d/(1.43\sqrt{2}\sqrt{m^{*}}\left(\left(1+y\right)^{2}+\frac{\left(1-y\right)^{2}}{\left(1+2z\right)^{1/2}}\right))$. Here $a=a_{0}d^{2}/(2\pi\sigma_x\sigma_y)$ and $a_{12}=a_{012}d^{2}/(2\pi\sigma_x\sigma_y)$ denote the lattice-projected effective scattering lengths. The coefficient of $n_0^2$ combines the positive bare mean field with the negative lattice contribution. When their sum is negative, the positive higher-order LHY term can stabilize a self-bound minimum. Dividing by the total density gives
\begin{equation}
	\frac{E_{g\text{3D}}}{N}=\frac{\pi\hbar^{2}}{ma^{2}}\left(\alpha x+lx^{3/2}\right),\label{EN0}
\end{equation}
where $\alpha=1+y-1.43\sqrt{2}\,\frac{a}{d}\sqrt{\frac{m^{*}}{m}}\left(\left(1+y\right)^{2}+\frac{\left(1-y\right)^{2}}{\left(1+g_{3}x\right)^{1/2}}\right)$ and $l=\frac{32\sqrt{2}}{15\sqrt{\pi}}\frac{m^{*}}{m}\left(\left(1+y\right)^{5/2}+\frac{\left(1-y\right)^{5/2}}{\left(1+g_{3}x\right)^{1/2}}\right)$.

As in standard droplet theory, a self-bound state is identified by a finite-density minimum of $E/N$. Figure~\ref{fig:3Ddroplet}(a) shows this minimum for $a/d=0.01$ and $t_{1,2}=5$, corresponding to $m^{*}/m\approx1.18$.
We set $y=-0.95$, so the bare mean-field coefficient is $1+y=0.05$ and both Bogoliubov branches are real in the parameter range considered. At $g_3=0$, the lattice contribution is $a/a_{\mathrm{cr}}\simeq0.08$ for the parameters of Fig.~\ref{fig:3Ddroplet}, leaving a weak net attraction. More generally, a self-bound minimum requires the coefficient of $n_0^2$ in Eq.~(\ref{Eg3D}) to be negative, that is $1+y<a/a_{\mathrm{cr}}$, in place of the free-space condition $y<-1$. At $a/d=0.01$ and $z=0$ this gives $y<-0.919$, so that $y=-0.95$ lies inside a window of width $0.081$, and at $a/d=0.02$ the window widens to $y<-0.849$. Reducing $a/d$ narrows the window instead: at $a/d\simeq0.006$ the threshold has moved to $y\simeq-0.950$, so this is the value at which the fixed working point $y=-0.95$ ceases to be bound, not the closure of the entire $y>-1$ window, which persists for any $a/d>0$. The thresholds quoted here are evaluated at $z=0$ and $m^{*}/m\approx1.18$. The lattice therefore does not simply displace the free-space droplet condition, it opens a window at $y>-1$ whose width is set by $a/d$. We vary $x=na^3$ at fixed $g_3$ and retain $2z=g_3x$ when differentiating $E/N$. For the plotted range $x\le10^{-5}$ and $g_3\le10^5$, one has $z\le0.5$; at the quoted equilibria, the largest value is approximately $0.10$. These values specify the $p$-wave window used in the calculation.

To identify the equilibrium density quantitatively, we evaluate $\partial\left(E/N\right)/\partial\left(na^{3}\right)$. Since $z=g_{3}x/2$ depends explicitly on density, we treat $g_{3}$ rather than $z$ as the density-independent control parameter, and obtain
\begin{align}
	\frac{\partial\left(E/N\right)}{\partial\left(na^{3}\right)}\Big/\frac{\pi\hbar^{2}}{ma^{2}} & =1+y-C_{2}P_{2}+\frac{C_{2}g_{3}x\left(1-y\right)^{2}}{2\left(1+g_{3}x\right)^{3/2}}\nonumber \\
	& \quad+\frac{3}{2}C_{5}x^{1/2}P_{5}-\frac{C_{5}g_{3}x^{3/2}\left(1-y\right)^{5/2}}{2\left(1+g_{3}x\right)^{3/2}},\label{partial3D}
\end{align}
where $C_{2}=1.43\sqrt{2}\,\frac{a}{d}\sqrt{\frac{m^{*}}{m}}$ and $C_{5}=\frac{32\sqrt{2}}{15\sqrt{\pi}}\frac{m^{*}}{m}$ are density-independent constants, while $P_{2}=\left(1+y\right)^{2}+\frac{\left(1-y\right)^{2}}{\sqrt{1+g_{3}x}}$ and $P_{5}=\left(1+y\right)^{5/2}+\frac{\left(1-y\right)^{5/2}}{\sqrt{1+g_{3}x}}$ carry the residual density dependence through $2z=g_{3}x$; the coefficients of Eq.~(\ref{EN0}) are thereby recovered as $\alpha=1+y-C_{2}P_{2}$ and $l=C_{5}P_{5}$.

The droplet density is then determined by the stationary condition
\begin{align*}
\frac{\partial\left(E/N\right)}{\partial\left(na^{3}\right)} & =0,
\end{align*}
which is equivalent to the zero-pressure condition~\cite{Petrov2015,Staudinger2018}. As shown in Fig.~\ref{fig:3Ddroplet}(b), the zero crossing of $\partial\left(E/N\right)/\partial\left(na^{3}\right)$ provides a direct numerical identification of the droplet density. One also sees that increasing $g_{3}$ shifts the zero crossing toward lower density, in agreement with the movement of the minimum in Fig.~\ref{fig:3Ddroplet}(a). Quantitatively, the equilibrium densities are $n_{\text{eq}}a^{3}=4.4$, $3.5$, $2.7$, and $2.0$ (in units of $10^{-6}$) for $g_{3}=0$, $2\times10^{4}$, $5\times10^{4}$, and $10^{5}$, corresponding to self-consistent $z_{\text{eq}}=g_{3}n_{\text{eq}}a^{3}/2=0$, $0.035$, $0.068$, and $0.10$.

The corresponding three-dimensional depletion function has the asymptotic form
\begin{align}
	h_{\text{3D}}\left(s\right) & =\frac{4\sqrt{2}}{3\pi s}\left(\frac{\left(1-y\right)^{3/2}}{\left(2z+1\right)^{1/2}}+\left(y+1\right)^{3/2}\right).\label{hs3D}
\end{align}

Substituting Eq.~(\ref{hs3D}) into the general expression for the quantum depletion, we obtain 
\begin{align}
	\frac{\Delta n_{\text{3D}}}{n} & =\frac{2\sqrt{2}}{3\pi^{1/2}}\frac{m^*}{m}\left(na^{3}\right)^{1/2}\left(\frac{\left(1-y\right)^{3/2}}{\left(2z+1\right)^{1/2}}+\left(y+1\right)^{3/2}\right),\label{n3D}
\end{align}
Equation~(\ref{n3D}) has the weak-coupling scaling $\Delta n/n\propto\sqrt{na^3}$~\cite{Lee1957,Andersen2004}. Its value at the equilibrium density tests the weak-depletion approximation for each working point.

\subsection{Quasi-1D asymptotic behavior\label{sec:1D}}
We next consider $s\ll1$, where the system is quasi-one-dimensional. The equation of state has the familiar low-dimensional form: a repulsive mean-field contribution and an attractive beyond-mean-field correction~\cite{Petrov2016,Hu2020}. Because the transverse bandwidth remains finite, the regime is quasi- rather than strictly one-dimensional~\cite{Du2023,Sinha2023}.

In the following quasi-1D analysis, $a$ and $a_{12}$
denote the one-dimensional scattering lengths defined by
$g^{\mathrm{1D}}=-2\hbar^{2}/(ma)$ and
$g_{12}^{\mathrm{1D}}=-2\hbar^{2}/(ma_{12})$,
with $g_{12}^{\mathrm{1D}}=g_{12}/d^{2}$.

In the limit $s\ll1$ the scaling function approaches the constant $f_{\text{1D}}\equiv\lim_{s\to0}f(s)$, given by
\begin{align}
f_{\text{1D}} & =\frac{4\sqrt{2}}{3}\frac{\left(1-y\right)^{3/2}+\left(1+y\right)^{3/2}\sqrt{1+2z}}{\sqrt{1+2z}}.\label{fs1D}
\end{align}

Substitution of Eq.~(\ref{fs1D}) into Eq.~(\ref{Eg}) gives
\begin{eqnarray}
&&\!\!\!\!\!\!\!\!\!\!\!\!\frac{E^{\text{1D}}_{\text{g}}}{L} =g^{\text{1D}}(n^\text{1D}_{0})^{2}+g^{\text{1D}}_{12}(n^\text{1D}_{0})^{2}\nonumber \\
&&\!\!\!\!\!\!\!\!\!\!\!\!\!-\frac{\sqrt{2}\left(g^{\text{1D}}n_{0}^{\text{1D}}\right)^{\frac{3}{2}}}{3\pi}\!\!\left(\frac{2m}{\hbar^{2}}\right)^{\frac{1}{2}}\!\!\frac{\left(1-y\right)^{\frac{3}{2}}\!+\!\left(1+y\right)^{\frac{3}{2}}\!\!(1+2z)^{\frac{1}{2}}}{(1+2z)^{\frac{1}{2}}},
\label{Eg1D}
\end{eqnarray}
Here $n_{0\text{1D}}=n_{0}d^{2}$ is the line density and $g^{\text{1D}}=g/d^{2}$ the effective one-dimensional coupling.

\begin{figure}[htbp]
 \begin{centering} 
  \includegraphics[scale=0.46]{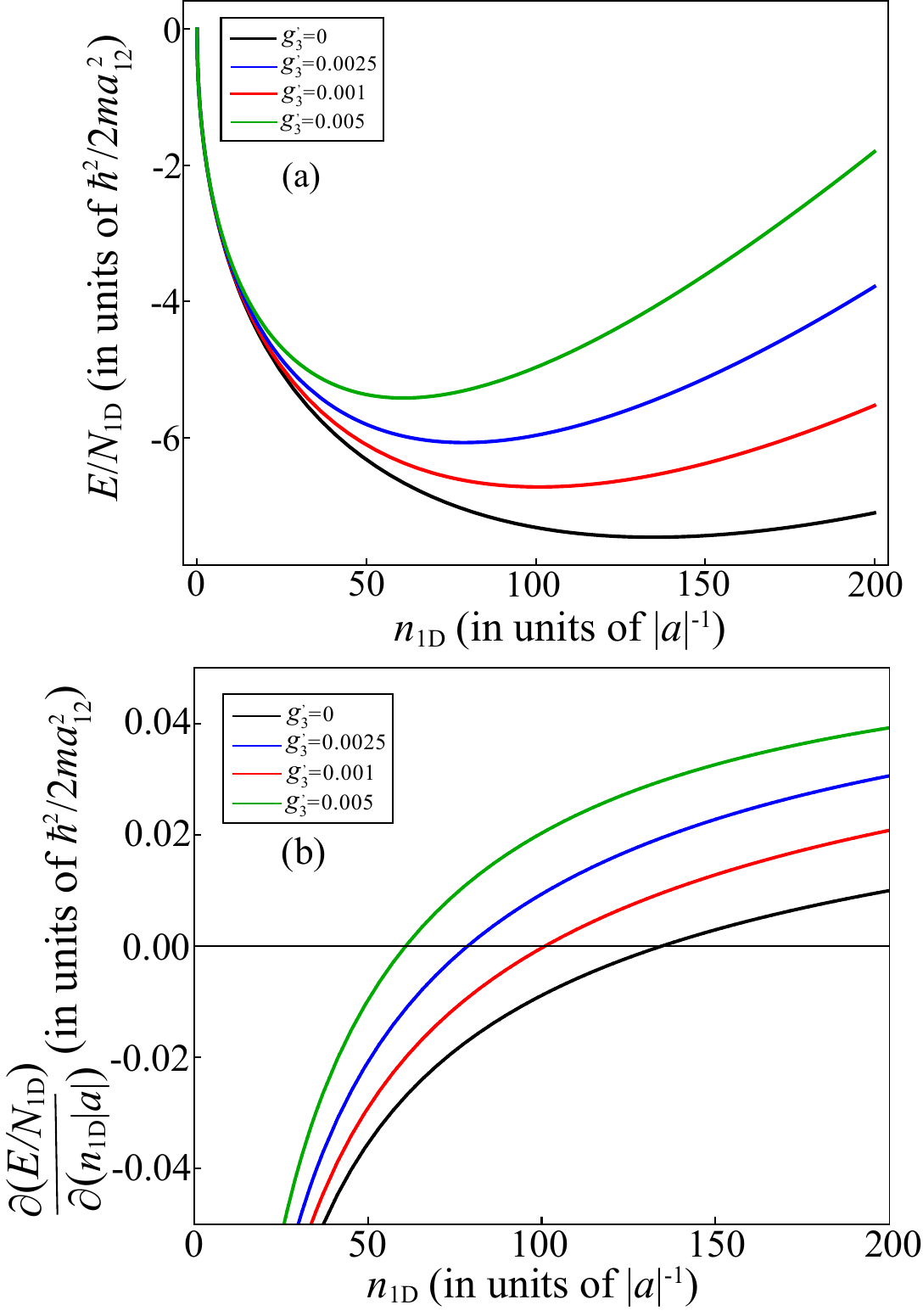} 
  \end{centering} 
 \caption{(a) Asymptotic quasi-1D energy per particle $E/N_{\text{1D}}$ (in units of $\hbar^{2}/2ma_{12}^{2}$) as a function of the dimensionless density $n_{\text{1D}}$ (in units of $|a|^{-1}$) at $y=-0.95$, for $g_{3}^{\prime}=0$ (black), $0.001$ (red), $0.0025$ (blue), and $0.005$ (green). A finite-density minimum appears, indicating the formation of a self-bound quasi-1D droplet.
(b) Corresponding derivative $\partial(E/N_{\text{1D}})/\partial(n_{\text{1D}}|a|)$ (same units), whose zero crossing determines the equilibrium droplet density. As $g_{3}^{\prime}$ increases, the zero crossing shifts toward lower density, showing that the $p$-wave interaction weakens the fluctuation-induced binding of the quasi-1D droplet.
    \label{fig:1Ddroplet}
}
\end{figure}

Dividing Eq.~(\ref{Eg1D}) by the total density gives the energy per particle,
\begin{align}
\frac{E_{\text{g1D}}}{N_{\text{1D}}} & =\frac{\hbar^{2}}{2ma^{2}_{12}}\rho\beta\left(\beta-1\right)\nonumber \\
 & -\frac{2\hbar^{2}}{3\pi ma^{2}_{12}}\sqrt{\rho\beta}\frac{\left(\beta+1\right)^{3/2}+\left(\beta-1\right)^{3/2}\sqrt{1+g^{\prime}_{3}\rho}}{\sqrt{1+g^{\prime}_{3}\rho}},
\end{align}
Here $\beta=-1/y=a_{12}/|a|$ and $\rho=n_{\mathrm{1D}}|a|$ uses the total line density $n_{\mathrm{1D}}=2n_{0\mathrm{1D}}$. We vary the density at fixed $g'_3=2z/\rho$, so $2z=g'_3\rho$. The positive mean-field term competes with attractive fluctuations, producing the minimum shown in Fig.~\ref{fig:1Ddroplet}(a).

To relate the two regimes quantitatively, write $a_{\perp}$ for the lattice-projected three-dimensional length of Sec.~\ref{sec:3D} and retain $a$ for the one-dimensional length used here. Matching $g^{\text{1D}}=-2\hbar^{2}/(ma)$ to $g^{\text{1D}}=g/d^{2}$ with $g=4\pi\hbar^{2}a_{\perp}/m$ gives $|a|=d^{2}/(2\pi a_{\perp})$, hence $na_{\perp}^{3}=2\pi\rho\,(a_{\perp}/d)^{4}$ and
\begin{equation}
g^{\prime}_{3}=2\pi\left(\frac{a_{\perp}}{d}\right)^{4}g_{3}.\label{g3conv}
\end{equation}
At $a_{\perp}/d=0.01$ this gives $g^{\prime}_{3}\simeq6.3\times10^{-8}g_{3}$, so the couplings $g^{\prime}_{3}=0.001$, $0.0025$, and $0.005$ of Fig.~\ref{fig:1Ddroplet} correspond to $g_{3}\simeq1.6\times10^{4}$, $4.0\times10^{4}$, and $8.0\times10^{4}$, comparable to the values used in Fig.~\ref{fig:3Ddroplet}.

We use the same interaction ratio $y=-0.95$ as in the three-dimensional calculation. At $g'_3=0$, the equilibrium line density $\rho_{\mathrm{eq}}\simeq135$ gives the per-component Lieb--Liniger parameter $\gamma_{\mathrm{LL}}=4/\rho_{\mathrm{eq}}\simeq0.030$. At the largest coupling shown, $\rho_{\mathrm{eq}}\simeq61$ gives $\gamma_{\mathrm{LL}}\simeq0.066$. These values lie in the weak-coupling regime used for the one-dimensional fluctuation energy~\cite{Petrov2016}.

Differentiating at fixed $g'_3$ gives
\begin{widetext}
\begin{align}
\frac{\partial\left(E_{\text{g1D}}/N_{\text{1D}}\right)}{\partial \rho}=\frac{\hbar^{2}}{2a^{2}_{12}m}\beta\left\{ \beta-1+\frac{2g^{\prime}_{3}\sqrt{\rho}(\beta+1)^{3/2}}{3\pi(g^{\prime}_{3}\rho+1)^{3/2}\sqrt{\beta}}-\frac{2(\beta-1)^{3/2}}{3\pi\sqrt{\rho\beta}}-\frac{2(\beta+1)^{3/2}}{3\pi\sqrt{g^{\prime}_{3}\rho+1}\sqrt{\rho\beta}}\right\} .\label{partial1D}
\end{align}
\end{widetext}

The droplet density is then determined by the stationary condition
\begin{equation}
\frac{\partial\left(E_{\text{g1D}}/N_{\text{1D}}\right)}{\partial \rho} =0.
\end{equation}
At the minima shown in Fig.~\ref{fig:1Ddroplet}, the derivative changes from negative to positive and $E/N_{\mathrm{1D}}<0$. These two conditions identify the locally stable self-bound solutions of the bulk equation of state.

The equilibrium density decreases from $\rho_{\mathrm{eq}}=135$ at $g'_3=0$ to $101$, $79$, and $61$ at $g'_3=0.001$, $0.0025$, and $0.005$, respectively. The corresponding self-consistent values of $z_{\mathrm{eq}}=g'_3\rho_{\mathrm{eq}}/2$ are approximately $0.05$, $0.10$, and $0.15$. Over the plotted interval $\rho\le200$, all curves have $z\le0.5$.

\subsection{Lattice-induced droplet binding across the crossover}

The two asymptotic regimes share a common structure. At $y=-0.95$ the bare mean field is repulsive in both limits, and the binding is supplied by an induced attraction. In the three-dimensional limit, the large-$s$ expansion of Eq.~(\ref{fs}) contributes a negative lattice term at order $n^{2}$, while the LHY term at order $n^{5/2}$ is positive. In the quasi-one-dimensional limit, the small-$s$ limit of the same scaling function gives the attractive fluctuation energy. The dominant attractive contribution therefore passes from the lower-order lattice term to the higher-order fluctuation term without a change in the sign of the mean-field coefficient. In both limits the factor $(1+2z)^{-1/2}$ multiplies the spin-branch contribution, while the density-branch contribution is unchanged.

This differs from the free-space three-dimensional droplet, for which self-binding requires $y<-1$~\cite{Petrov2015}. At $y=-0.95$ the free-space mean field is repulsive and no such droplet forms; in the present model the lattice term supplies the attraction instead. Working at $y>-1$ also has a technical consequence: by Eq.~(\ref{reality}) both Bogoliubov branches are real at all momenta throughout the scan, so the prescription of evaluating the mean-field and LHY contributions at split values of $y$~\cite{Petrov2015,Hu2020} is not required, and the factor $(1+2z)^{-1/2}$ need not be assigned to one order or the other. As noted after Eq.~(\ref{reality}), the reality condition is itself independent of the lattice parameters: the lattice moves the droplet window onto the side where no imaginary branch occurs, rather than rendering an otherwise imaginary branch real.

\begin{figure}[htbp]
 \begin{centering}
  \includegraphics[scale=0.46]{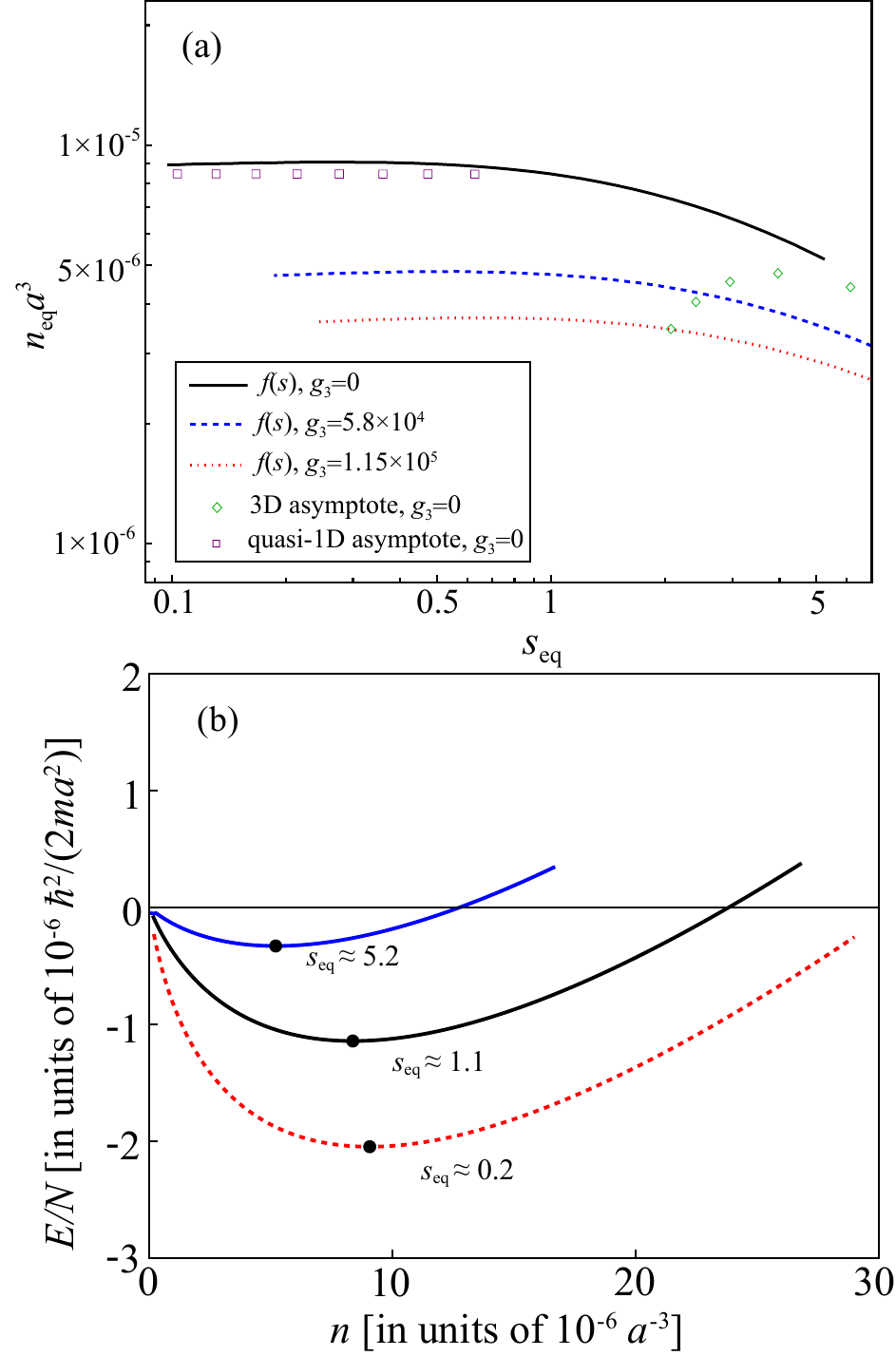}
  \end{centering}
 \caption{(a) Equilibrium density $n_{\mathrm{eq}}a^3$ versus $s_{\mathrm{eq}}$ at $y=-0.95$ and $a/d=0.01$, calculated from the full $f(s)$ [Eq.~(\ref{fs})]. Curves are for $g_3=0$ (black solid), $5.76\times10^{4}$ (blue dashed), and $1.15\times10^{5}$ (red dotted). They are labelled by the density-independent $g_3$ because $z_{\mathrm{eq}}=g_3n_{\mathrm{eq}}a^3/2$ varies along each branch; over the range shown the self-consistent values are $z_{\mathrm{eq}}=0.075$--$0.140$ and $0.113$--$0.212$. Open diamonds and squares denote the 3D [Eq.~(\ref{fs3D})] and quasi-1D [Eq.~(\ref{fs1D})] asymptotes at $g_3=0$, plotted for $s_{\mathrm{eq}}\gtrsim2$ and $s_{\mathrm{eq}}\lesssim0.7$, respectively. (b) Energy per particle in units of $10^{-6}\hbar^2/(2ma^2)$, the same units as Fig.~\ref{fig:3Ddroplet}, versus density in units of $10^{-6}a^{-3}$, at $g_3=0$. The three lattice depths give $s_{\mathrm{eq}}\simeq5.2$ (blue), $1.1$ (black), and $0.2$ (red dashed). Dots mark the negative-energy minima.
    \label{fig:crossover}
}
\end{figure}

In the intermediate regime neither asymptotic equation of state is quantitatively sufficient, so we evaluate the full $f(s)$. Figure~\ref{fig:crossover}(a) shows the equilibrium density at $y=-0.95$ and $a/d=0.01$. At $g_{3}=0$ the three-dimensional form, Eq.~(\ref{fs3D}), underestimates the equilibrium density by $16\%$ at $s_{\mathrm{eq}}\simeq5.2$, by $35\%$ at $s_{\mathrm{eq}}\simeq2.2$ and by $49\%$ at $s_{\mathrm{eq}}\simeq1.3$, and reaches the $10\%$ level only near $s_{\mathrm{eq}}\simeq7$; the quasi-one-dimensional form, Eq.~(\ref{fs1D}), overestimates it by $10\%$ at $s_{\mathrm{eq}}\simeq1.7$ and remains within $7\%$ for $s_{\mathrm{eq}}\lesssim1.3$. Neither limit is accurate to $10\%$ in the range $1.7\lesssim s_{\mathrm{eq}}\lesssim7$, which delineates the transdimensional window according to this criterion for these parameters. The deviations are amplified by the near cancellation in the coefficient of $n^{2}$: at $s_{\mathrm{eq}}\simeq5.2$ the scaling function differs from its three-dimensional asymptote by less than $2\%$, while the equilibrium density differs by $16\%$. For $g_3=0$, the calculated branch approaches the three-dimensional and quasi-one-dimensional asymptotes at the corresponding ends and differs from both in the intermediate range. The branch is continuous over the range shown: within this equation of state, the lattice-bound three-dimensional state and the fluctuation-bound quasi-one-dimensional state are connected without an intervening unbinding transition.

The curves at $g_3=5.76\times10^{4}$ and $1.15\times10^{5}$ lie below the $g_3=0$ result over the range shown. Because $z_{\mathrm{eq}}=g_3n_{\mathrm{eq}}a^3/2$ follows the equilibrium density, it is not constant along a branch: the self-consistent values span $0.075$--$0.140$ and $0.113$--$0.212$, respectively. At $g_3=0$, Fig.~\ref{fig:crossover}(b) shows finite-density minima with $E/N<0$ at three representative tunnelling strengths, $s_{\mathrm{eq}}\simeq5.2$, $1.1$, and $0.2$. These label the tunnelling strengths sampled and not the three regimes defined by the $10\%$ criterion above: $s_{\mathrm{eq}}\simeq5.2$ is the large-tunnelling end of the present scan and still lies inside the transdimensional window, while at $s_{\mathrm{eq}}\simeq1.1$ the quasi-one-dimensional form is already accurate to better than $10\%$.

Increasing the $p$-wave coupling lowers the equilibrium density and reduces the binding energy in magnitude, in both limits and in between. In the three-dimensional asymptotic calculation the density decreases by approximately $21\%$, $38\%$, and $54\%$ at $g_3=2\times10^{4}$, $5\times10^{4}$, and $10^{5}$; the corresponding quasi-one-dimensional decreases are $25\%$, $42\%$, and $55\%$ at $g'_3=0.001$, $0.0025$, and $0.005$. The common sign of these shifts follows from the factor $(1+2z)^{-1/2}$, which multiplies the entire spin-branch contribution. In the three-dimensional expansion, Eq.~(\ref{Eg3D}), this factor modifies both the attractive $n^{2}$ term and the repulsive $n^{5/2}$ term, and the direction of the equilibrium-density shift follows from their combined contribution to the equation of state. The shift is therefore consistent with binding supplied by an induced attraction acting on the spin branch rather than on the mean-field background, but the sign of the shift does not by itself identify the source of the attraction. Nor does it distinguish the two regimes, which differ in the order in density at which the attraction enters rather than in the sign of the shift.

At the three equilibrium points quoted in Fig.~\ref{fig:crossover}(b), Eq.~(\ref{n}) gives depletion fractions of $0.35\%$, $0.88\%$, and $2.0\%$ for $s_{\mathrm{eq}}=5.2$, $1.1$, and $0.2$, respectively; the value reaches approximately $2.6\%$ at $s_{\mathrm{eq}}=0.1$. At the three-dimensional equilibria of Fig.~\ref{fig:3Ddroplet}, the depletion decreases from $0.36\%$ at $g_3=0$ to $0.22\%$ at $g_3=10^5$. These finite-tunnelling values support the weak-depletion approximation at the sampled points~\cite{Xu2006,Li2022}.

At fixed finite $g_3$, one has $2z=g_3x\to0$ as $x=na^3\to0$. In the three-dimensional equation of state, $(E/N)/(\pi\hbar^2/ma^2)=\alpha_0x+O(x^{3/2})$, where $\alpha_0=1+y-C_2[(1+y)^2+(1-y)^2]$ and $C_2$ is defined after Eq.~(\ref{partial3D}). For the parameters of Fig.~\ref{fig:3Ddroplet}, $\alpha_0\simeq-0.034$. The low-density slope therefore does not change sign at any finite $g_3$ in this model: the weakening of the binding is continuous rather than terminating at a finite critical coupling, and the equilibrium density and binding energy approach zero continuously. Two limitations should be stated. First, $\alpha_0$ follows from a near cancellation between $1+y=0.05$ and $a/a_{\mathrm{cr}}\simeq0.083$, so its magnitude, and with it the two-digit density shifts quoted above, is sensitive to the accuracy of the lattice coefficient of Eq.~(\ref{lattconst}). Second, the statement concerns the bulk equation of state and does not by itself establish that a finite droplet survives, which would additionally require the surface energy at these densities.

\section{Conclusion\label{sec:conclusion}}

We calculated the bulk equation of state and the quantum depletion of a symmetric Bose mixture with interspecies $p$-wave interactions in a transverse optical lattice, using a coherent-state path integral at Gaussian order~\cite{Salasnich2016}, together with the three-dimensional and quasi-one-dimensional limits of the resulting scaling functions. The expressions reduce to the known lattice and free-space results when the $p$-wave coupling is removed~\cite{Orso2006,Li2022,Petrov2015,Petrov2016}.

At $y=-0.95$ the calculation connects a lattice-bound three-dimensional state to a quasi-one-dimensional state bound by attractive fluctuations, and the full crossover calculation is required between these limits~\cite{Ilg2018}. In both limits the bare mean field is repulsive, so the attraction is induced rather than intrinsic: it passes from the lower-order lattice term to the higher-order fluctuation term as the transverse tunnelling is reduced. Because the interspecies $p$-wave coupling enters only the spin branch, through the factor $(1+2z)^{-1/2}$ that multiplies the attractive and the repulsive spin-branch terms alike, increasing it lowers the equilibrium density and reduces the binding energy in magnitude over the whole range studied. The shift is consistent with binding supplied by an induced attraction rather than by the mean-field background, but does not on its own identify that attraction. Because its sign is the same in both limits, it also does not discriminate between the two mechanisms, which differ in the order in density at which the attraction enters.

As an estimate of the energy scales, take $m=39u$ and $d=532\,\mathrm{nm}$~\cite{DErrico2007,Tanzi2018}, giving $E_R/h=4.52\,\mathrm{kHz}$. For the projected length $a/d=0.01$ and a per-component density $n_0=1.73\times10^{13}\,\mathrm{cm}^{-3}$~\cite{Catani2008,Gadway2010}, one obtains $gn_0/h\simeq300\,\mathrm{Hz}$. Using $J_i/E_R\simeq4t_i^{3/4}e^{-2\sqrt{t_i}}/\sqrt\pi$~\cite{Bloch2008,Li2022}, symmetric depths $t_i=5$ and $13$ give $J_i/E_R\simeq0.086$ and $0.011$, respectively. At this fixed density, they correspond to $s\simeq5.2$ and $0.69$. These estimates specify tunnelling and interaction scales; the equilibrium scan in Fig.~\ref{fig:crossover} also includes the variation of density with lattice depth.

The same parameters fix the binding energies. At $a/d=0.01$ one has $\pi\hbar^{2}/(ma^{2})=h\times28.8\,\mathrm{MHz}$, so the minima of Fig.~\ref{fig:crossover}(b) correspond to $|E|/N=h\times1.5$, $5.2$ and $9.5\,\mathrm{Hz}$, that is $72$, $250$ and $454\,\mathrm{pK}$, at $s_{\mathrm{eq}}\simeq5.2$, $1.1$ and $0.2$. The binding deepens monotonically towards the quasi-one-dimensional end. These energies define characteristic binding temperature scales $T_{b}=|E|/(Nk_{B})$; a quantitative criterion for thermal stability requires a finite-temperature treatment. Increasing the projected length changes both the dimensionless equation of state and the equilibrium density. Although the energy unit $\pi\hbar^{2}/(ma^{2})$ decreases as $a$ increases, the recalculated minima are deeper: at $a/d=0.02$ the same calculation gives $h\times25$ to $45\,\mathrm{Hz}$, that is $1.2$ to $2.2\,\mathrm{nK}$, at the cost of a stronger requirement on the band gap. At the three-dimensional end of this range $na^{3}\simeq1.4\times10^{-4}$ and the quantum depletion is about $2\%$; as at $a/d=0.01$, it grows towards the quasi-one-dimensional end. Within this model a realization further requires an interspecies $p$-wave resonance compatible with these $s$-wave parameters~\cite{Dong2016,Wang2013,Venu2023}, together with the band-gap, effective-range and lifetime conditions noted below; these conditions have to be met jointly.

The length $a$ in this estimate is the lattice-projected length. At $t_1=t_2=5$, the projection factor $d^2/(2\pi\sigma_x\sigma_y)\simeq2.8$ gives a free-space length $a_0\simeq0.0036d\simeq36a_{\mathrm B}$, where $a_{\mathrm B}$ is the Bohr radius. Holding $a/d$ fixed as the lattice depth changes requires adjusting $a_0$. A specific experimental realization also requires compatible intra- and interspecies scattering parameters~\cite{Thalhammer2008,Tanzi2018} and an assessment of effective-range corrections and losses near the $p$-wave resonance~\cite{Idziaszek2009,Venu2023}.

Unequal masses or intraspecies couplings would mix the density and spin modes, and species-dependent Wannier orbitals would change the cancellation of the transverse $p$-wave vertex~\cite{Soltan-Panahi2011}; extending the projection to that case would test the range of applicability of the present axial-vertex model. Beyond the bulk equation of state, the surface energy~\cite{Petrov2015}, collective modes~\cite{Orso2006,Hu2011}, and the dynamics of the crossover itself~\cite{Cavicchioli2025} remain open, and are required to convert the equilibrium densities reported here into predictions for finite droplets.

\begin{acknowledgments}
We thank Tao Yu, Kangkang Li, Ying Hu and Biao Wu for stimulating discussions and helpful suggestions. This work was supported by the National Natural Science Foundation of China under Grant No. 12574301, the Zhejiang Provincial Natural Science Foundation of China under Grant No. LZ25A040004 and the Key Project of the National Natural Science Foundation of China Joint Funds under Grant No. U25A20197.
\end{acknowledgments}
\section*{Appendix}

\appendix

\section{Detailed derivation of Eq.~(\ref{Eg})}\label{A}

In this Appendix, we derive the analytical expression of the scaling function $f(s)$ defined in Eq.~(\ref{fs}). The starting point is the zero-temperature Gaussian grand potential, whose contribution to the ground-state energy density takes the form
\begin{align}
\frac{E_{\text{g}}}{L^{3}} & =gn^{2}_{0}+g_{12}n^{2}_{0}-(I_{+}+I_{-}),\label{Eg-1}
\end{align}
where $I_{+}$ and $I_{-}$ denote the contributions of the density and spin branches, respectively. At the mean-field saddle point these are $I_{\sigma}=(2V)^{-1}\sum_{\boldsymbol k}(\mathcal A_{\boldsymbol k,\sigma}-E_{\boldsymbol k,\sigma})$, so that $-(I_{+}+I_{-})$ is the zero-temperature Gaussian contribution per unit volume of Eq.~(\ref{Omegag}).
\begin{eqnarray}
I_{+} & =&\frac{\left(gn_{0}\right)^{3/2}}{8\pi^{3}d^{2}}\left(\frac{2m}{\hbar^{2}}\right)^{1/2}\int^{\pi}_{-\pi}d^{2}k\int^{\infty}_{0}dk_{z}^{\prime}\nonumber \\
 && \left[k^{\prime2}_{z}+\left(1+s\gamma+y\right)-\sqrt{\left(k^{\prime2}_{z}+s\gamma\right)\left(k^{\prime2}_{z}+s\gamma+2+2y\right)}\right]\nonumber \\
 && \overset{k^{\prime2}_{z}+s\gamma=\zeta}{=}\frac{\left(gn_{0}\right)^{3/2}}{16\pi^{3}d^{2}}\left(\frac{2m}{\hbar^{2}}\right)^{1/2}\int^{\pi}_{-\pi}d^{2}k\int^{\infty}_{s\gamma}d\zeta\frac{1}{\sqrt{\zeta-s\gamma}}\nonumber \\
 && \times\left\{\zeta+1+y-\sqrt{\zeta\left[\zeta+2\left(1+y\right)\right]}\right\}\nonumber \\
 && \overset{\frac{s\gamma}{\zeta}=\tau}{=}\frac{\left(gn_{0}\right)^{3/2}}{16\pi^{3}d^{2}}\left(\frac{2m}{\hbar^{2}}\right)^{1/2}\int^{\pi}_{-\pi}d^{2}k\int^{1}_{0}\frac{s\gamma}{\tau^{2}}d\tau\frac{1}{\sqrt{s\gamma}\sqrt{\frac{1}{\tau}-1}}\nonumber \\
 && \times\left\{\frac{s\gamma}{\tau}+\left(1+y\right)-\sqrt{\frac{s\gamma}{\tau}\left[\frac{s\gamma}{\tau}+2\left(1+y\right)\right]}\right\}\nonumber \\
 & =&\frac{\left(gn_{0}\right)^{3/2}}{4\pi d^{2}}\left(\frac{2m}{\hbar^{2}}\right)^{1/2}\int^{\pi}_{-\pi}d^{2}k\frac{\sqrt{s\gamma}}{6\pi}\nonumber \\
 && \times\left\{\left(1+y+s\gamma\right){}_{2}F_{1}\left[-\frac{1}{2},\frac{1}{2},1,\frac{-2\left(1+y\right)}{s\gamma}\right]\right.\nonumber \\
 && \left.-\left(2+2y+s\gamma\right){}_{2}F_{1}\left[\frac{1}{2},\frac{1}{2},1,\frac{-2\left(1+y\right)}{s\gamma}\right]\right\},
\end{eqnarray}
and 
\begin{eqnarray}
I_{-} & =&\frac{\left(gn_{0}\right)^{3/2}}{8\pi^{3}d^{2}}\left(\frac{2m}{\hbar^{2}}\right)^{1/2}\int^{\pi}_{-\pi}d^{2}k\int^{\infty}_{0}dk^{\prime}_{z}\nonumber \\
 && \times\left\{\left(1+2z\right)k^{\prime2}_{z}+\left(1+s\gamma-y\right)\right.\nonumber \\
 && \left.-\sqrt{\left[k^{\prime2}_{z}\left(1+2z\right)+s\gamma\right]\left[k^{\prime2}_{z}\left(1+2z\right)+s\gamma+2\left(1-y\right)\right]}\right\}\nonumber \\
 && \overset{(1+2z)k^{\prime2}_{z}+s\gamma=\eta}{=}\frac{\left(gn_{0}\right)^{3/2}}{16\pi^{3}d^{2}}\left(\frac{2m}{\hbar^{2}}\right)^{1/2}\int^{\pi}_{-\pi}d^{2}k\int^{\infty}_{s\gamma}d\eta\nonumber \\
 && \frac{1}{\sqrt{1+2z}\sqrt{\eta-s\gamma}}\left\{\eta+1-y-\sqrt{\eta\left[\eta+2-2y\right]}\right\}\nonumber \\
 && \overset{\frac{s\gamma}{\eta}=j}{=}-\frac{\left(gn_{0}\right)^{3/2}}{16\pi^{3}d^{2}}\left(\frac{2m}{\hbar^{2}}\right)^{1/2}\int^{\pi}_{-\pi}d^{2}k\int^{0}_{1}\frac{s\gamma}{j^{2}}dj\nonumber \\
 && \frac{1}{\sqrt{1+2z}\sqrt{\frac{s\gamma}{j}-s\gamma}}\left\{\frac{s\gamma}{j}+1-y-\sqrt{\frac{s\gamma}{j}\left[\frac{s\gamma}{j}+2-2y\right]}\right\}\nonumber \\
 && =\frac{\left(gn_{0}\right)^{3/2}}{16\pi^{3}d^{2}}\left(\frac{2m}{\hbar^{2}}\right)^{1/2}\int^{\pi}_{-\pi}\frac{1}{\sqrt{\left(1+2z\right)s\gamma}}d^{2}k\int^{1}_{0}\frac{1}{\sqrt{1-j}}\frac{1}{\sqrt{j}}dj\nonumber \\
 && \times\left\{\frac{(s\gamma)^{2}}{j^{2}}+\frac{s\gamma}{j}\left(1-y\right)-\frac{(s\gamma)^{2}}{j^{2}}\sqrt{\left[1+\frac{\left(2-2y\right)j}{s\gamma}\right]}\right\}\nonumber \\
 && =\frac{\left(gn_{0}\right)^{3/2}}{4\pi d^{2}}\left(\frac{2m}{\hbar^{2}}\right)^{1/2}\int^{\pi}_{-\pi}\frac{1}{\sqrt{1+2z}}d^{2}k\frac{\sqrt{s\gamma}}{6\pi}\nonumber \\
 && \times\left\{\left(1+s\gamma-y\right){}_{2}F_{1}\left[-\frac{1}{2},\frac{1}{2},1,\frac{2\left(y-1\right)}{s\gamma}\right]\right.\nonumber \\
 && \left.-\left(2+s\gamma-2y\right){}_{2}F_{1}\left[\frac{1}{2},\frac{1}{2},1,\frac{2\left(y-1\right)}{s\gamma}\right]\right\}.
\end{eqnarray}

Combining the above results, we find that $I_{+}+I_{-}=\frac{\left(gn_{0}\right)^{3/2}}{4\pi d^{2}}\left(\frac{2m}{\hbar^{2}}\right)^{1/2}f(s)$. Substituting these expressions back into the energy density yields Eqs.~(\ref{Eg}) and~(\ref{fs}).
\bibliography{xyref}
\end{document}